\documentclass{JHEP3}
\usepackage{amssymb}

\usepackage{amsmath,amssymb,graphics}
\usepackage{epsfig,multicol}
\usepackage{graphicx}
\usepackage{epsfig,subfigure}

\newcommand {\nn}    {\nonumber}

\title{Bulk Matters on Symmetric and Asymmetric de Sitter Thick Branes}

\author{Yu-Xiao Liu$^1$\footnote{Corresponding author.},
        Zhen-Hua Zhao$^{2,1}$, Shao-Wen Wei$^1$, Yi-Shi Duan$^1$ \\
  $^1$Institute of Theoretical Physics,
  Lanzhou University, Lanzhou 730000, P. R. China\\
  $^2$Institute of Modern Physics,
  Chinese Academy of Sciences,
  Lanzhou 730000, P. R. China\\
  E-mail: \email{liuyx@lzu.edu.cn}, \email{zhaozhenhua@impcas.ac.cn},
          \email{weishaow06@lzu.cn}, \email{ysduan@lzu.edu.cn}}

\abstract{ An asymmetric thick domain wall solution with de Sitter
($dS$) expansion in five dimensions can be constructed from a
symmetric one by using a same scalar (kink) with different
potentials. In this paper, by presenting the mass-independent
potentials of Kaluza--Klein (KK) modes in the corresponding
Schr\"{o}dinger equations, we investigate the localization and
mass spectra of various bulk matter fields on the symmetric and
asymmetric $dS$ thick branes. For spin 0 scalars and spin 1
vectors, the potentials of KK modes in the corresponding
Schr\"{o}dinger equations are the modified P\"{o}schl-Teller
potentials, and there exist a mass gap and a series of continuous
spectrum. It is shown that the spectrum of scalar KK modes on the
symmetric $dS$ brane contains only one bound mode (the massless
mode). However, for the asymmetric $dS$ brane with a large
asymmetric factor, there are two bound scalar KK modes: a zero
mode and a massive mode. For spin 1 vectors, the spectra of KK
modes on both $dS$ branes consist of a bound massless mode and a
set of continuous ones, i.e., the asymmetric factor does not
change the number of the bound vector KK modes. For spin 1/2
fermions, two types of kink-fermion couplings are investigated in
detail. For the usual Yukawa coupling $\eta\bar{\Psi}\phi\Psi$,
there exists no mass gap but a continuous gapless spectrum of KK
states. For the scalar-fermion coupling $\eta\bar{\Psi}\sin(
\frac{\phi}{\phi_0})\cos^{-\delta}(\frac{\phi}{\phi_0})\Psi$ with
a positive coupling constant $\eta$, there exist some discrete
bound KK modes and a series of continuous ones. The total number
of bound states increases with the coupling constant $\eta$. For
the case of the symmetric $dS$ brane and positive $\eta$, there
are $N_L(N_L\geq 1)$ left chiral fermion bound states (including
zero mode and massive KK modes) and $N_L-1$ right chiral fermion
bound states (including only massive KK modes). For the asymmetric
$dS$ brane scenario, the asymmetric factor $a$ reduces the number
of the bound fermion KK modes. For large enough $a$, there would
not be any right chiral fermion bound mode, but at least one left
chiral fermion zero mode.}


\begin{document}

\section{Introduction}

The idea of embedding our universe in a higher dimensional space
has received a great of renewed attention. The suggestion that
extra dimensions may not be compact
\cite{RubakovPLB1983136,VisserPLB1985,Randjbar-DaemiPLB1986,rs,Lykken}
or large \cite{AntoniadisPLB1990,ADD} can provide new insights for
solving gauge hierarchy problem \cite{ADD}, i.e., the large
difference in magnitude between the Planck and electroweak scales,
and the long-standing cosmological constant problem
\cite{RubakovPLB1983136,Randjbar-DaemiPLB1986,CosmConst}.
According to the brane scenarios, gravity is free to propagate in
all dimensions, while all the matter fields (electromagnetic,
Yang-Mills etc.) are confined to a 3--brane in a high-dimensional
space. In Ref. \cite{rs}, an alternative scenario of the
compactification had been proposed. In this scenario, the internal
manifold does not need to be compactified to the Planck scale any
more, it can be large, or even infinite non-compact, which is one
of reasons why this new compactification scenario has attracted so
much attention. Among all of the brane world models, there is an
interesting and important model in which extra dimensions comprise
a compact hyperbolic manifold \cite{StojkovicCHM}. The model is
known to be free of usual problems that plague the original ADD
models and share many common features with Randall-Sundrum (RS)
models.

Recently, an increasing interest has been focused on the study of
thick brane scenario in higher dimensional space-time
\cite{dewolfe,gremm,Csaki,CamposPRL2002,WangPRD2002,varios,ThickBrane},
since in more realistic models the thickness of the brane should be
taken into account. A virtue of these models is that the branes can
be obtained naturally rather than introduced by hand. In this
scenario the scalar field configuration is usually a kink, which
provides a thick brane realization of the brane world as a domain
wall in the bulk. However, the inclusion of the gravitational
evolution into a dynamic thick wall is a highly non-trivial problem
because of the non-linearity of the Einstein equations. For this
reason, there are not so many analytic solutions of a dynamic thick
domain wall. The symmetric de Sitter ($dS$) branes have been studied
in five and higher dimensional spacetimes, for examples in
\cite{WangPRD2002,MinamNPB737,SasakuraJHEP2002}. Ref.
\cite{asymdSBrane2} presented a method to construct asymmetric thick
$dS$ brane solutions from known ones, where the spacetimes
associated to them are physically different. With the method,
asymmetric brane worlds with $dS$ expansion were obtained. These
branes interpolate between two spacetimes with different
cosmological constants, and the vacua correspond to $dS$ and $AdS$
geometry. It was shown that gravity is localized on such branes.

In brane world scenarios, an important and complex question is
localization of various bulk fields on a brane by a natural
mechanism. It is well known that massless scalar fields
\cite{BajcPLB2000} and graviton \cite{rs} can be localized on
branes of different types. However, spin 1 Abelian vector fields
can not be localized on the RS brane in five dimensions, but can
be localized on the RS brane in some higher-dimensional cases
\cite{OdaPLB2000113} or on the thick $dS$ brane and Weyl thick
brane \cite{Liu0708}. The localization problem of spin 1/2
fermions on thick branes is interesting and important. Fermions do
not have normalizable zero modes in five and six dimensions
without the scalar-fermion coupling
\cite{BajcPLB2000,OdaPLB2000113,Liu0708,NonLocalizedFermion,Volkas0705.1584,IchinosePRD2002,Ringeval,GherghettaPRL2000,Neupane,RandjbarPLB2000,KoleyCQG2005,DubovskyPRD2000,0803.1458}.
In Ref. \cite{KoleyCQG2005}, the authors obtained trapped discrete
massive fermion states on the brane, which in fact are quasi-bound
and have a finite probability of escaping into the bulk. In fact,
fermions can escape into the bulk by tunnelling, and the rate
depends on the parameters of the scalar potential
\cite{DubovskyPRD2000}. In five dimensions, with the
scalar--fermion coupling, there may exist a single bound state and
a continuous gapless spectrum of massive fermion Kaluza--Klein
(KK) states \cite{Liu0708,ThickBraneWeyl}. While in some other
brane models, there exist finite discrete KK states (mass gap) and
a continuous gapless spectrum starting at a positive $m^2$
\cite{ThickBrane4,Liu0803}.

Since a physically different asymmetric thick $dS$ brane solution
can be constructed from a known symmetric one by including an
asymmetric factor, we will address the localization and mass
spectrum problems of various bulk matters on the symmetric and
asymmetric $dS$ branes, and investigate the influence of the
asymmetric factor on the mass spectra of bulk matters in this
paper. We will show that all bulk matters (scalars, vectors and
fermions) can be localized on these branes and the corresponding
mass spectra have a mass gap (for spin 1/2 fermions the
scalar-fermion coupling should not be the usual Yukawa coupling
$\eta\bar{\Psi}\phi \Psi$ in order to trap the zero mode). The
large asymmetric factor increases the number of the scalar bound
states but reduces that of the fermion ones, and does not change
the number of the vector bound states.

The organization of the paper is as follows: In section
\ref{SecModel}, we first review the symmetric and asymmetric $dS$
thick branes in 5-dimensional space-time. Then, in section
\ref{SecLocalize}, we study the localization and mass spectra of
various bulk fields on the symmetric and asymmetric thick branes
by presenting the shapes of the potentials of the corresponding
Schr\"{o}dinger problem. For spin 1/2 fermions, we consider two
different types of scalar-fermion interactions. Finally, the
conclusion and summary are given.

\section{Review of the symmetric and asymmetric thick branes}
\label{SecModel}

Let us consider thick branes arising from a real scalar field $\phi$
with a scalar potential $V(\phi)$. The action for such a system is
given by
\begin{equation}
S = \int d^5 x \sqrt{-g}\left [ \frac{1}{2\kappa_5^2} R-\frac{1}{2}
g^{MN}\partial_M \phi
\partial_N \phi - V(\phi) \right ],
\label{action}
\end{equation}
where $R$ is the scalar curvature and $\kappa_5^2=8 \pi G_5$ with
$G_5$ the 5-dimensional Newton constant. Here we set $\kappa_5=1$.
The line-element for a 5-dimensional spacetime with planar-paralell
symmetry is assumed as
\begin{eqnarray}
 ds^2&=&\text{e}^{2A(z)}\big(\hat{g}_{\mu\nu}(x)dx^\mu dx^\nu
          + dz^2\big)    \nonumber \\
 &=&\text{e}^{2A(z)}\big(-dt^2+e^{2\beta t}dx^i dx^i + dz^2\big),
\label{linee}
\end{eqnarray}
where $\text{e}^{2A(z)}$ is the warp factor and $z$ stands for the
extra coordinate. For the positive constant $\beta>0$, we will
have dynamic solutions. The scalar field is considered to be a
function of $z$ only, i.e., $\phi=\phi(z)$. In the model, the
potential could provide a realization of a thick brane, and the
soliton configuration of the scalar field dynamically generate the
domain wall configuration with warped geometry. The field
equations generated from the action (\ref{action}) with the ansatz
(\ref{linee}) reduce to the following coupled nonlinear
differential equations
\begin{eqnarray}
\phi'^2 & = & 3(A'^2-A''-\beta^2), \\
V(\phi) & = & \frac{3}{2} e^{-2A}
 (-3A'^2-A''+3\beta^2),\\
\frac{dV(\phi)}{d\phi} &  = & e^{-2A}(3A'\phi'+\phi''),
\end{eqnarray}
where the prime denotes derivative with respect to $z$. For positive
and vanishing ${\beta}$ we will obtain dynamic and static solutions,
respectively.



A symmetric thick domain wall with $dS$ expansion in five
dimensions for the potential
\begin{eqnarray}
V(\phi)=\frac{1+3\delta}{2\delta}\ 3\beta^{2}\left(\cos
\frac{\phi}{\phi_{0}} \right)^{2(1-\delta)},
          \label{potencial goetz}
\end{eqnarray}
was found in Refs. \cite{Goetz:1990,Gass:1999gk}:
\begin{eqnarray}
 e^{2A}&=&\cosh^{-2\delta}\left(\frac{\beta z}{\delta}\right) ,
         \label{e2A1} \\
 \phi~~&=&\phi_{0}\arctan\left(\sinh \frac{\beta z}{\delta}\right),
         \label{phi1}
\end{eqnarray}
where $\phi_{0} =\sqrt{3\delta(1-\delta)}$, $0<\delta<1$ and
$\beta>0$. In this system, The scalar field takes values
$\pm\phi_0\pi/2$ at $z\rightarrow\pm \infty$, corresponding to two
consecutive minima of the potential with cosmological constant
$\Lambda=0$. The scalar configuration in fact is a kink, which
provides a thick brane realization of the brane world as a domain
wall in the bulk. $\delta$ plays the role of the wall's thickness.
The thick brane has a well-defined distributional thin wall limit
when $\delta\rightarrow 0$ \cite{Guerrero:2002ki} and can localize
gravity on the wall \cite{WangPRD2002}. Note that for the cases
where $1/2 < \delta < 1$, the hypersurfaces $|z|=\infty$ represent
non-scalar spacetime singularities \cite{WangPRD2002}.

\begin{figure}[htb]
\begin{center}
\includegraphics[width=7cm,height=5cm]{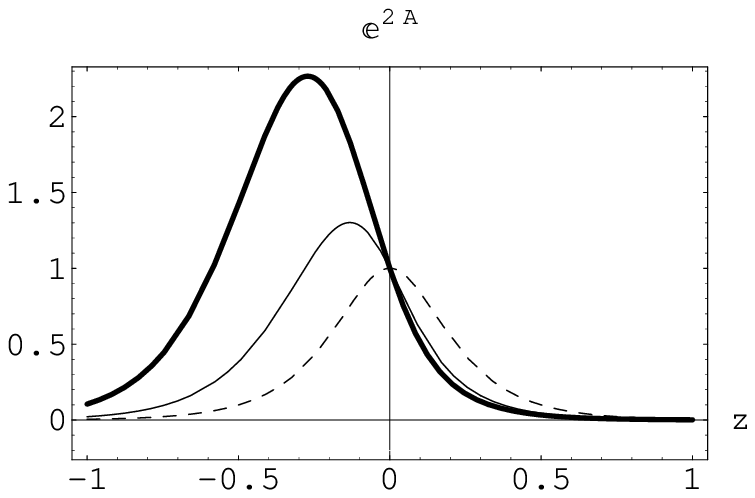}
\includegraphics[width=7cm,height=5cm]{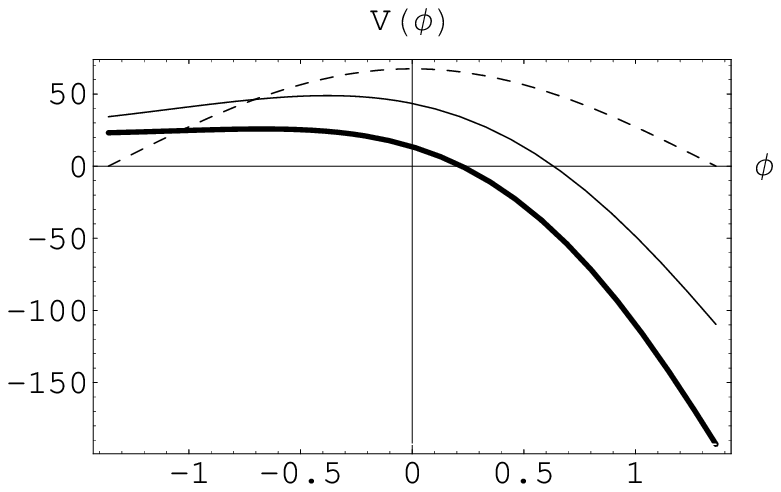}
\includegraphics[width=7cm,height=5cm]{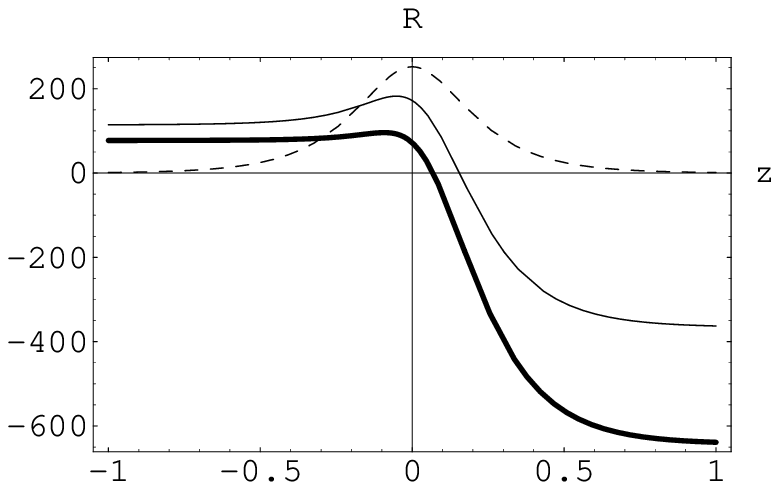}
\includegraphics[width=7cm,height=5cm]{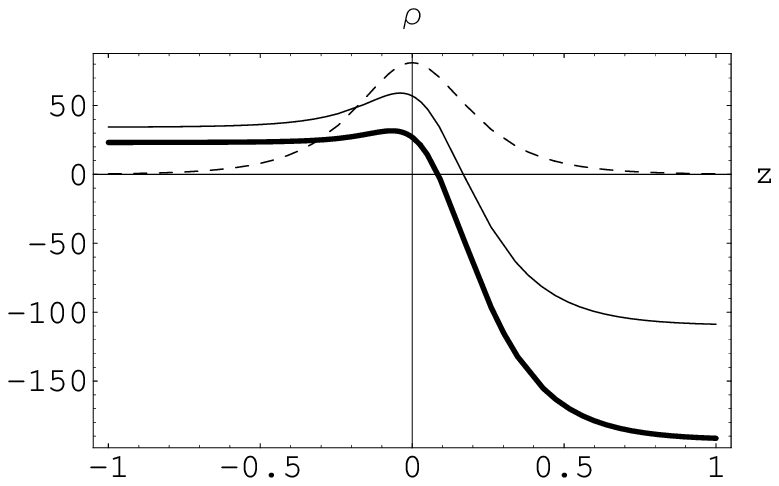}
\end{center}
\caption{The shapes of the metric factor $e^{2A}$, scalar
potential $V(\phi)$, the scalar curvature $R$, and the energy
density $\rho$ for the $dS$ branes with $\beta=3$. The parameter
$a$ is set to $a=0$ for dashed lines, $a=2$ for thin lines, and
$a=3$ for thick lines.}
 \label{fig_dSBrane}
\end{figure}

An asymmetric thick domain wall solution with $dS$ expansion in
five dimensions for the same kink configuration $\phi$ in
(\ref{phi1}) was found in Ref. \cite{asymdSBrane2}:
\begin{eqnarray}
e^{-A}&=&\cosh^{\delta}\left(\frac{\beta z}{\delta}\right)
  + \frac{i a\delta}{\beta - 2\beta\delta}
      \cosh^{-\delta}\left(\frac{\beta z}{\delta}\right)
      \coth\left(\frac{\beta z}{\delta}\right)\nonumber \\
  && \times  \left|\sinh\left(\frac{{\beta}z}{\delta}\right)\right| ~
     {}_2F_1\left(\frac{1}{2}-\delta,\frac{1}{2},\frac{3}{2}-\delta,
            \cosh^2\left(\frac{{\beta}z}{\delta}\right)\right),
  \label{newgeneral}
\end{eqnarray}
where ${}_{2}F_{1}$ is the hypergeometric function. Here we will
consider the case $\delta=1/2$ for convenience:
\begin{eqnarray}
e^{2A}&=&\frac{\beta^2 \text{sech} 2\beta z}
       {[\beta+a\arctan(\tanh \beta z)]^2},
       \qquad   \label{e2A2} \\
\phi~~ &=&\frac{\sqrt{3}}{2}\arctan\left(\sinh 2\beta z\right),
       \label{phi2} \\
V(\phi)&=&
  \frac{3}{2}\left|{\cos\frac{2\phi}{\sqrt{3}}}\right|
  \bigg\{-4a^2 + 5\beta^2
    -8a\beta \tan \frac{2\phi}{\sqrt{3}} \nonumber \\
 && +5 a^2\arctan^2\left[\tanh\left(\frac{1}{2}
    \text{arcsinh}\left(\tan\frac{2\phi}{\sqrt{3}}
    \right)\right)\right]\nonumber \\
 && +2a\arctan\left[\tanh\left(\frac{1}{2}
    \text{arcsinh}\left(\tan\frac{2\phi}{\sqrt{3}}
    \right)\right)\right]
    \left(5\beta-4a\tan\frac{2\phi}{\sqrt{3}}\right)
    \bigg\}, \label{Vphi2}
\end{eqnarray}
where $|a|<4\beta/\pi$ in order to prevent singularities in the
metric tensor. The parameter $a$ decides the asymmetry of the
solution. For $a=0$, we recover the symmetric domain wall
solution. For positive (negative) $a$, the spacetime for
$z\rightarrow +\infty$ is asymptotically $AdS$ ($dS$) with
cosmological constant $-3a(4\beta+a\pi)$ and for $z\rightarrow
-\infty\ $ is asymptotically $dS$ ($AdS$) with cosmological
constant $3a(4\beta-a\pi)$. The scalar curvature $R$ and the
energy density $\rho$ for the $dS$ brane are calculated as
follows:
\begin{eqnarray}
 R &=& 4 \text{sech} 2\beta z  \left[7\beta^2-5 a^2
      +14 a \beta \arctan (\tanh \beta z)\right.  \nonumber \\
 &&\left.+7 a^2 \arctan^2(\tanh\beta z)
      -10 a (\beta +a \arctan(\tanh\beta z))\sinh 2\beta z \right],
       \label{R2} \\
 \rho &=& \text{sech} 2\beta z \left[3\beta^2-2a^2
      +6 a \beta \arctan (\tanh \beta z)\right.  \nonumber \\
 &&\left.+3 a^2 \arctan^2(\tanh\beta z)
      -4 a (\beta +a \arctan(\tanh\beta z))\sinh 2\beta z \right].
       \label{EnergyDensity2}
\end{eqnarray}
The shapes for the metric factor $e^{2A}$, the potential $V(\phi)$,
the scalar curvature $R$, and the density energy $\rho$ are shown in
Fig. \ref{fig_dSBrane}.

\section{Localization and mass gaps of various matters on the thick branes}
\label{SecLocalize}

In this section let us investigate whether various bulk mater
fields such as spin 0 scalars, spin 1 vectors and spin 1/2
fermions can be localized on the thick branes by means of only the
gravitational interaction. Of course, we have implicitly assumed
that various bulk fields considered below make little contribution
to the bulk energy so that the solutions given in previous section
remain valid even in the presence of bulk fields. We will analyze
the spectra of various mater fields for the thick brane by
presenting the potential of the corresponding Schr\"{o}dinger
equation. It can be seen from the following calculations that the
mass-independent potential can be obtained conveniently with the
conformally flat metric (\ref{linee}).

\subsection{Spin 0 scalar fields}
We first study localization of a real scalar field on the branes
obtained in previous section. Let us start by considering the
action of a massless real scalar coupled to gravity
\begin{eqnarray}
S_0 = - \frac{1}{2} \int d^5 x  \sqrt{-g}\; g^{M N}
\partial_M \Phi \partial_N \Phi.
\label{scalarAction}
\end{eqnarray}
By considering the conformally flat metric  (\ref{linee}) the
equation of motion derived from (\ref{scalarAction}) is read
\begin{eqnarray}
 \frac{1}{\sqrt{-\hat{g}}}\partial_\mu(\sqrt{-\hat{g}}
   \hat{g}^{\mu \nu}\partial_\nu \Phi)
 +e^{-3A} \partial_z
  \left(e^{3A}\partial_z \Phi \right) = 0. \label{scalarEOM}
\end{eqnarray}

Then, by decomposing $\Phi(x,z) =   \sum_n
\phi_n(x)\chi_n(z)e^{3A/2}$ and demanding $\phi_n(x)$ satisfies the
4-dimensional massive Klein--Gordon equation
$\left(\frac{1}{\sqrt{-\hat{g}}}\partial_\mu(\sqrt{-\hat{g}}
   \hat{g}^{\mu \nu}\partial_\nu) -\mu_n^2 \right)\phi_n(x)=0 $, we
obtain the equation for $\chi_n(z)$:
\begin{eqnarray}
  \left[-\partial^2_z+ V_0(z)\right]{\chi}_n(z)
  =\mu_n^2 {\chi}_n(z),
  \label{SchEqScalar1}
\end{eqnarray}
which is a Schr\"{o}dinger equation with the effective potential
given by
\begin{eqnarray}
  V_0(z)=\frac{3}{2} A'' + \frac{9}{4}A'^{2}, \label{VScalar}
\end{eqnarray}
where $\mu_n$ is the mass of the KK excitations. It is clear that
$V_0(z)$ defined in (\ref{VScalar}) is a mass-independent
potential.

The full 5-dimensional action (\ref{scalarAction}) reduces to the
standard 4-dimensional action for the massive scalars
\begin{eqnarray}
 S_0&=& - \frac{1}{2} \sum_{n}\int d^4 x \sqrt{-\hat{g}}
     \bigg(\hat{g}^{\mu\nu}\partial_\mu\phi_{n}
           \partial_\nu\phi_{n}
           +\mu_{n}^2 \phi^2_{n}
     \bigg), \label{ScalarEffectiveAction}
\end{eqnarray}
when integrated over the extra dimension, in which it is required
that Eq. (\ref{SchEqScalar1}) is satisfied and the following
orthonormality condition is obeyed:
\begin{eqnarray}
 \int^{\infty}_{-\infty} dz
 \;\chi_m(z)\chi_n(z)=\delta_{mn}.
 \label{normalizationCondition1}
\end{eqnarray}

For the symmetric and asymmetric $dS$ brane world solutions
(\ref{e2A1}) and (\ref{e2A2}), the potentials corresponding to
(\ref{VScalar}) are
\begin{eqnarray}
 V_0^S(z)&=& \frac{3\beta^2}{4\delta}
  \left(3\delta-(2+3\delta){\text{sech}^2 (\beta z/\delta)} \right)
   \label{VScalarSymmetricdS}
\end{eqnarray}
and
\begin{eqnarray}
 V_0^A(z)&=& \frac{9 \beta ^2}{4}
   +\frac{15 a \beta ^2 \text{sech} 2\beta z  \tanh 2\beta z }
         {2(\beta +a \arctan\tanh \beta z)} \nonumber \\
   &&-\frac{3\beta^2\left(7\beta^2-5a^2 +7a\arctan\tanh\beta z
            (2\beta +a\arctan \tanh \beta z )\right)}
         {4 (\beta +a\arctan\tanh\beta z)^2\cosh^2 (2\beta
         z)},\;\;  (\delta=\frac{1}{2})\quad
   \label{VScalardS}
\end{eqnarray}
respectively. For the case $a=0$, the potential (\ref{VScalardS})
is reduced to (\ref{VScalarSymmetricdS}) with $\delta=1/2$:
\begin{eqnarray}
 V_0^S(z)&=& \frac{3}{4}\beta^2
  \left(3-7{\text{sech}^2 (2\beta z)} \right).
  \;\;  (\delta=\frac{1}{2})
   \label{VScalarSymmetricdS2}
\end{eqnarray}

\begin{figure}[htb]
\begin{center}
\includegraphics[width=7cm,height=5cm]{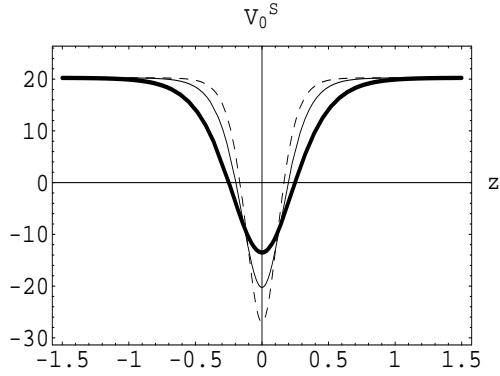}
\end{center}
\caption{The shapes of the potential $V_0^S(z)$ for the symmetric
$dS$ brane. The parameters are set to $\beta=3$, $\delta=1/2$ for
dashed line, $\delta=2/3$ for thin line, and $\delta \rightarrow
1$ for thick line.}
 \label{fig_V0S_dS_Brane}
\end{figure}

We first investigate the potential (\ref{VScalarSymmetricdS}) for
the symmetric $dS$ brane. It has a minimum (negative value)
$-\frac{3\beta^2}{2\delta}$ at $z=0$ and a maximum (positive
value) $\frac{9}{4}\beta^2$ at $z=\pm\infty$. Let $p=\beta/\delta$
and $q=1+3\delta/2$, Eq. (\ref{SchEqScalar1}) with the potential
(\ref{VScalarSymmetricdS}) turns into the well-known
Schr\"{o}dinger equation with
$E_n=\mu_n^2-\frac{9}{4}\delta^2p^2$:
\begin{eqnarray}
\Bigl[-\partial_z^2  - q(q-1)p^2{\rm sech}^2(pz)\Bigr]~ \chi_n =
E_n ~ \chi_n . \label{SchEqScalar2}
\end{eqnarray}
For this equation with a modified P\"{o}schl-Teller potential, the
energy spectrum of bound states is found to be
\begin{eqnarray}
E_n=-p^2(q-1-n)^2
\end{eqnarray}
or
\begin{eqnarray}
\mu_n^2=n(3\delta-n)\frac{\beta^2}{\delta^2},
\label{spectrumScalar2}
\end{eqnarray}
where $n$ is an integer and satisfies $0\leq n
<\frac{3}{2}\delta$. It is clear that the energy for $n = 0$ or
$\mu_0=0$ always belongs to the spectrum of the potential
(\ref{VScalarSymmetricdS}) for $\delta>0$. For
$0<{\delta}{\leq}\frac{2}{3}$, there is only one bound state,
i.e., the ground state
\begin{eqnarray}
\chi_0(z)=\sqrt{\frac{\beta\Gamma(\frac{1}{2}+\frac{3\delta}{2})}
             { \delta\sqrt{\pi}\;  \Gamma(\frac{3\delta}{2}) }}
  {\rm sech}^{3\delta/2}({\beta}z/\delta) \label{groundScalar1}
\end{eqnarray}
with $\mu_0=0$, which is just the normalized zero-mass mode and
also shows that there is no tachyonic scalar mode. The continuous
spectrum starts with $\mu^2 = \frac{9}{4}\beta^2$ and
asymptotically turn into plane waves, which represent delocalized
KK massive scalars. For $\frac{2}{3}<{\delta}<1$, there are two
bound states, one is the ground state (\ref{groundScalar1}),
another is the first exited state
\begin{eqnarray}
\chi_1(z)\propto
  {\rm sech}^{3\delta/2}({\beta}z/\delta) \sinh z \label{exitedScalar1}
\end{eqnarray}
with mass $\mu_1^2=(3\delta-1){\beta^2}/{\delta^2}$. The
continuous spectrum also start with $\mu^2 = \frac{9}{4}\beta^2$.
From above analysis, we come to the conclusion: for
$0<{\delta}{\leq}\frac{1}{2}$, there is only one bound state (is
massless mode) for the symmetric potential
(\ref{VScalarSymmetricdS}).

\begin{figure}[htb]
\begin{center}
 \subfigure[$\delta=1/2$]  {\label{fig_V0S_Eigenvalue_Scalar_a}
  \includegraphics[width=7cm,height=5cm]{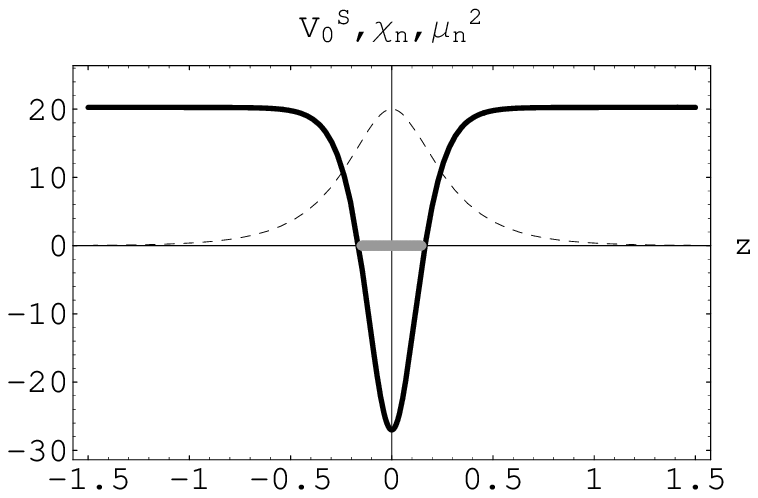}}
 \subfigure[$\delta=0.9$]  {\label{fig_V0S_Eigenvalue_Scalar_b}
  \includegraphics[width=7cm,height=5cm]{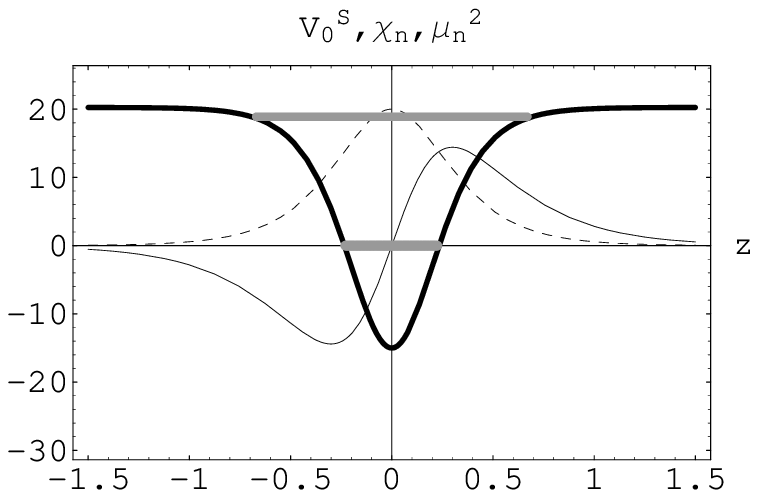}}
\end{center}
\caption{The shapes of the potential $V_0^S(z)$ (thick lines), KK
modes $\chi_n(z)$ (dashed lines for $\chi_0(z)$ and thin lines for
$\chi_1(z)$) and the mass spectrum (thick gray lines) for
symmetric $dS$ brane with $\beta=3$, $\delta=1/2$ and
$\delta=0.9$.}
 \label{fig_V0S_Eigenvalue_Scalar}
\end{figure}

\begin{figure}[htb]
\begin{center}
\includegraphics[width=7cm,height=5cm]{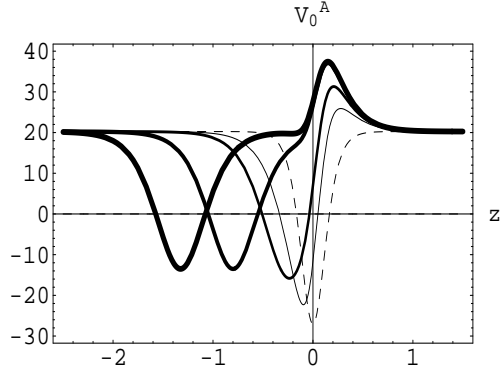}
\end{center}
\caption{The shapes of the potential $V_0^A(z)$ for the asymmetric
$dS$ brane. The parameters are set to $\beta=3$, $\delta=1/2$,
$a=0$ for dashed line, and $a=2$, 3, 3.78, 3.818 for solid lines
with thickness increases with $a$.}
 \label{fig_V0_dS_Brane}
\end{figure}

\begin{figure}[htb]
\begin{center}
 \subfigure[$a=2$]{\label{fig_V0_Eigenvalue_Scalar_a}
  \includegraphics[width=7cm,height=5cm]{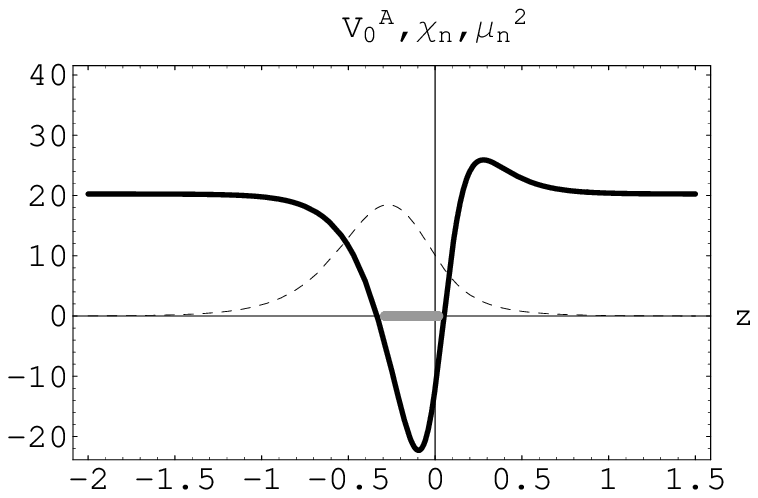}}
 \subfigure[$a=3.818$]
  {\label{fig_V0_Eigenvalue_Scalar_b}
  \includegraphics[width=7cm,height=5cm]{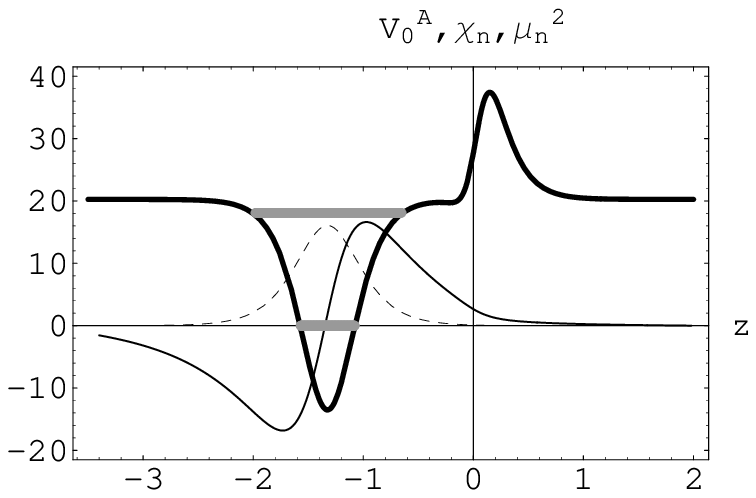}}
\end{center}
\caption{The shapes of the potential $V_0^A(z)$ (thick lines), KK
modes $\chi_n$ (dashed lines and thin lines) and the mass spectrum
$\mu_n^2$ (thick gray lines) for $dS$ brane with $\beta=3$,
$\delta=1/2$, and $a=2$, 3.818.}
 \label{fig_V0_Eigenvalue_Scalar}
\end{figure}

Next we turn to the potential (\ref{VScalardS}) for asymmetric
$dS$ brane. It has a negative value at some $z_0$ ($z_0<0$ for
$a>0$ and $z_0>0$ for $a<0$) and the asymptotic behavior:
$V_0^A(z=\pm \infty)=\frac{9}{4}\beta^2$, which implies that there
is also a mass gap. For the massless mode $\chi_0(z)$ with $\mu^2
= 0$, the Schr\"{o}dinger equation (\ref{SchEqScalar1}) with the
potential (\ref{VScalardS}) can be solved analytically, and the
normalizable eigenfunction is found to be
\begin{eqnarray}
\chi_0(z)\propto
  \left(\frac{\beta^2 \text{sech} 2\beta z}
       {[\beta+a\arctan(\tanh \beta z)]^2}\right)^{{3}/{4}}.
       \label{groundScalar2}
\end{eqnarray}
This zero mode is the ground state since it has no node. For the
limit $a \rightarrow 0$, the massless mode (\ref{groundScalar2})
is reduced to (\ref{groundScalar1}) but with $\delta=1/2$. Now, we
ask an interesting question: are there other bound states except
the zero mode for the asymmetric potential (\ref{VScalardS})? This
is very important for producing 4-dimensional massive scalars. If
the answer is yes, we will get massive scalars on the asymmetric
$dS$ brane. We have known that there is no any massive bound state
for the symmetric potential (\ref{VScalarSymmetricdS2}), the limit
case of the current asymmetric one. Hence we can extrapolate that
the answer should be no for small asymmetric factor $a$. However,
what will happen for large $a$? We note that the asymmetric
potential (\ref{VScalardS}) has the same asymptotic behavior as
the symmetric case: $V_0^A(\pm\infty)=\frac{9}{4}\beta^2$, but a
different minimum $V^A_{0\;min}$, which is larger than that of the
symmetric potential. The absolute value of the minimum of the
asymmetric potential decreases with the increase of the asymmetry
(see Fig. \ref{fig_V0_dS_Brane}). This leads to the increase of
the relative depth of the potential well
$V^A_0(\pm\infty)/|V^A_{0\;min}|$, which indicates that the
potential well may trap more bound stats. By numerical method, we
do get a massive bound state with $\mu_1^2=18.11$ at $a=3.818$
(see Fig. \ref{fig_V0_Eigenvalue_Scalar}).

\subsection{Spin 1 vector fields}

Next we turn to spin 1 vector fields. We begin with the 5D action
of a vector field
\begin{eqnarray}
S_1 = - \frac{1}{4} \int d^5 x \sqrt{-g} g^{M N} g^{R S} F_{MR}
F_{NS}, \label{actionVector}
\end{eqnarray}
where $F_{MN} = \partial_M A_N - \partial_N A_M$ as usual. From this
action and the background geometry (\ref{linee}), the equations of
motion $\frac{1}{\sqrt{-g}} \partial_M (\sqrt{-g} g^{M N} g^{R S}
F_{NS}) = 0$ are reduced to
\begin{eqnarray}
 \frac{1}{\sqrt{-\hat{g}}}\partial_\nu (\sqrt{-\hat{g}} ~
      \hat{g}^{\nu \rho}\hat{g}^{\mu\lambda}F_{\rho\lambda})
    +{\hat{g}^{\mu\lambda}}e^{-A}\partial_z
      \left(e^{A} F_{4\lambda}\right)  &=& 0, \\
 \partial_\mu (\sqrt{-\hat{g}}~ \hat{g}^{\mu \nu} F_{\nu 4}) &=& 0.
\end{eqnarray}

We assume that $A_4$ is $Z_2$-odd with respect to the extra
dimension $z$, which results in that $A_4$ has no zero mode in the
effective 4D theory. Furthermore, in order to consistent with the
gauge invariant equation $\oint dz A_4=0$, we use gauge freedom to
choose $A_4=0$. Under the assumption, the action
(\ref{actionVector}) is reduced to
\begin{eqnarray}
S_1 = - \frac{1}{4} \int d^5 x \sqrt{-g} \bigg\{
        g^{\mu\alpha} g^{\nu\beta} F_{\mu\nu}F_{\alpha\beta}
       +2e^{-A} g^{\mu\nu} \partial_z A_{\mu} \partial_z A_{\nu}
       \bigg\}.
\label{actionVector2}
\end{eqnarray}
Then, with the decomposition of the vector field
$A_{\mu}(x,z)=\sum_n a^{(n)}_\mu(x)\rho_n(z)e^{A/2}$, and importing
the orthonormality condition
\begin{eqnarray}
 \int^{\infty}_{-\infty} dz \;\rho_m(z)\rho_n(z)=\delta_{mn},
 \label{normalizationCondition2}
\end{eqnarray}
the action (\ref{actionVector2}) is read
\begin{eqnarray}
S_1 = \sum_{n}\int d^4 x \sqrt{-\hat{g}}~
       \bigg( - \frac{1}{4}\hat{g}^{\mu\alpha} \hat{g}^{\nu\beta}
             f^{(n)}_{\mu\nu}f^{(n)}_{\alpha\beta}
         - \frac{1}{2}\mu^2_{n} ~\hat{g}^{\mu\nu}
           a^{(n)}_{\mu}a^{(n)}_{\nu}
       \bigg),
\label{actionVector3}
\end{eqnarray}
where $f^{(n)}_{\mu\nu} = \partial_\mu a^{(n)}_\nu - \partial_\nu
a^{(n)}_\mu$ is the 4-dimensional field strength tensor, and it has
been required that the $\rho_n(z)$ satisfies the following
Schr\"{o}dinger equation
\begin{eqnarray}
  \left[-\partial^2_z +V_1(z) \right]{\rho}_n(z)=\mu_n^2
  {\rho}_n(z),  \label{SchEqVector1}
\end{eqnarray}
where the mass-independent potential is given by
\begin{eqnarray}
 V_1^S(z)&=& \frac{\beta^2}{4\delta}
  \left(\delta-(2+\delta){\text{sech}^2 (\beta z/\delta)} \right)
   \label{VVectorSymmetricdS}
\end{eqnarray}
and
\begin{eqnarray}
 V_1^A(z)&=& \frac{\beta ^2}{4}
   +\frac{3 a \beta ^2 \text{sech} 2\beta z  \tanh 2\beta z }
         {2(\beta +a \arctan\tanh \beta z)} \nonumber \\
   &&-\frac{\beta^2\left(5\beta^2-3a^2 +5a
            (2\beta +a\arctan \tanh \beta z )\arctan\tanh\beta z\right)}
         {4 (\beta +a\arctan\tanh\beta z)^2\cosh^2 (2\beta z)},
         \;\;  (\delta=\frac{1}{2})\quad
   \label{VVectordS}
\end{eqnarray}
for the symmetric and asymmetric $dS$ brane world solutions given
in previous section, respectively. The asymmetric potential at the
limit $a\rightarrow0$ (\ref{VVectordS}) is reduced to the
symmetric one (\ref{VVectorSymmetricdS}) with $\delta=1/2$:
\begin{eqnarray}
 V_1^S(z)&=& \frac{\beta^2}{4}
  \left(1-5{\text{sech}^2 (2\beta z)} \right).
   \label{VVectorSymmetricdS2}
\end{eqnarray}

The symmetric potential (\ref{VVectorSymmetricdS}) for arbitrary
$0<\delta<1$ has a minimum $-\frac{\beta^2}{2\delta}$ at $z=0$ and
a maximum $\beta^2/4$ at $z=\pm\infty$. Eq. (\ref{SchEqVector1})
with this potential can be turned into the following
Schr\"{o}dinger equation with a modified P\"{o}schl-Teller
potential:
\begin{eqnarray}
\Bigl[-\partial_z^2  - q(q-1)p^2{\rm sech}^2(pz)\Bigr]~ \chi_n =
E_n ~ \chi_n,  \label{SchEqVector2}
\end{eqnarray}
where $p=\beta/\delta$, $q=1+\delta/2$ and
$E_n=\mu_n^2-\frac{1}{4}\delta^2p^2$.  The energy spectrum of
bound states is found to be $E_n=-p^2(q-1-n)$ or
\begin{eqnarray}
 \mu_n^2=n(\delta-n)\frac{\beta^2}{\delta^2},\quad
  n \in \mathbb{Z},~ 0\leq n <\frac{1}{2}\delta.
 \label{spectrumVector2}
\end{eqnarray}
For $0<{\delta}<1$, we get only one bound state, i.e., the
normalized zero mode
\begin{eqnarray}
\rho_0(z)=\sqrt{\frac{\beta\Gamma(\frac{1}{2}+\frac{\delta}{2})}
             { \delta\sqrt{\pi}\;  \Gamma(\frac{\delta}{2}) }}
  {\rm sech}^{\delta/2}({\beta}z/\delta) \label{groundVector1}
\end{eqnarray}
with $\mu_0=0$. There is a mass gap between the zero mode and the
first excited mode. The continuous spectrum starts with $\mu^2 =
\frac{1}{4}\beta^2$ and asymptotically turn into plane waves,
which represent delocalized KK massive vectors.

\begin{figure}[htb]
\begin{center}
\includegraphics[width=7cm,height=5cm]{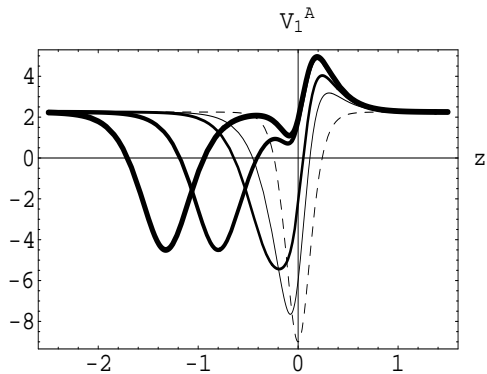}
\end{center}
\caption{The shapes of the potential $V_1^A(z)$ for the asymmetric
$dS$ brane. The parameters are set to $\beta=3$, $\delta=1/2$,
$a=0$ for dashed line, and $a=2$, 3, 3.78, 3.818 for solid lines
with thickness increases with $a$.}
 \label{fig_V1_dS_Brane}
\end{figure}

For the asymmetric $dS$ brane, the asymmetric potential
(\ref{VVectordS}) has a negative minimum value at some $z_0$ and
the asymptotic behavior: $V_1^A(z=\pm \infty)=\frac{1}{4}\beta^2$,
which implies that there is also a mass gap. The normalizable
massless mode $\rho_0(z)$ with $\mu^2 = 0$ is found to be
\begin{eqnarray}
\rho_0(z)\propto
  \left(\frac{\beta^2 \text{sech} 2\beta z}
       {[\beta+a\arctan(\tanh \beta z)]^2}\right)^{{1}/{4}}.
       \label{groundVector2}
\end{eqnarray}
This zero mode is the ground state since it has no node. For the
limit $a \rightarrow 0$, the massless mode (\ref{groundVector2})
is reduced to (\ref{groundVector1}) but with $\delta=1/2$. Now, we
also ask the question: are there other bound states except the
zero mode? Since there is no massive bound state for the symmetric
potential (\ref{VVectorSymmetricdS2}), we can conclude that the
answer is also no for small asymmetric factor $a$. Just as the
case of scalar, the absolute value of the minimum asymmetric
potential decreases with the increase of the asymmetry (see Fig.
\ref{fig_V1_dS_Brane}). However, we do not find massive bound
states by numerical method even for large $a$.

\begin{figure}[htb]
\begin{center}
\includegraphics[width=7cm,height=5cm]{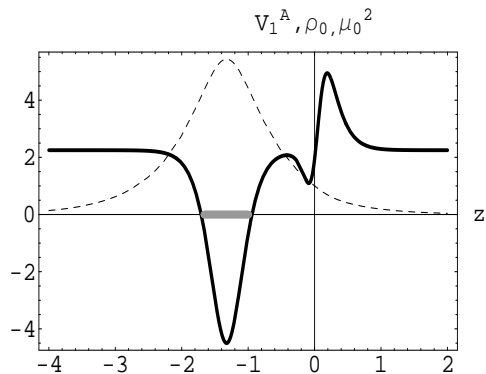}
\end{center}
\caption{The shapes of the potential $V_1^A(z)$ (thick line), zero
mode (dashed line) and the mass spectrum (thick gray line) for
$dS$ brane with $\beta=3$, $\delta=1/2$ and $a=3.818$.}
 \label{fig_V1_Eigenvalue_Scalar}
\end{figure}

It was shown in the RS model in $AdS_5$ space that a spin 1 vector
field is not localized neither on a brane with positive tension
nor on a brane with negative tension so the Dvali-Shifman
mechanism \cite{DvaliPLB1997} must be considered for the vector
field localization \cite{BajcPLB2000}. Here, it is turned out that
a vector field can be localized on the $dS$ thick branes and we do
not need to introduce additional mechanism for the vector field
localization in the case at hand. For $0<\delta<1$, we get only
one bound state which is the massless mode. Furthermore, there
exists a mass gap between the bound ground state and the first
exited state.

\subsection{Spin 1/2 fermion fields}

In five dimensions, fermions are four component spinors and their
Dirac structure is described by $\Gamma^M= e^M_{~\bar{M}}
\Gamma^{\bar{M}}$ with $e^M_{~\bar{M}}$ being the vielbein and
$\{\Gamma^M,\Gamma^N\}=2g^{MN}$. In this paper, $\bar{M}, \bar{N},
\cdots =0,1,2,3,5$ and $\bar{\mu}, \bar{\nu}, \cdots =0,1,2,3$
denote the 5D and 4D local Lorentz indices respectively, and
$\Gamma^{\bar{M}}$ are the flat gamma matrices in five dimensions.
In our set-up, the vielbein is given by
\begin{eqnarray}
e_M ^{~~\bar{M}}= \left(%
\begin{array}{ccc}
  \text{e}^{A} \hat{e}_\mu^{~\bar{\nu}} & 0  \\
  0 & \text{e}^{A}  \\
\end{array}%
\right),\label{vielbein_e}
\end{eqnarray}
$\Gamma^M=\text{e}^{-A}(\hat{e}^{\mu}_{~\bar{\nu}}
\gamma^{\bar{\nu}},\gamma^5)=\text{e}^{-A}(\gamma^{\mu},\gamma^5)$,
where $\gamma^{\mu}=\hat{e}^{\mu}_{~\bar{\nu}}\gamma^{\bar{\nu}}$,
$\gamma^{\bar{\nu}}$ and $\gamma^5$ are the usual flat gamma
matrices in the 4D Dirac representation. The Dirac action of a
massless spin 1/2 fermion coupled to the scalar is
\begin{eqnarray}
S_{1/2} = \int d^5 x \sqrt{-g} \left(\bar{\Psi} \Gamma^M
          (\partial_M+\omega_M) \Psi
          -\eta \bar{\Psi} F(\phi)\Psi\right), \label{DiracAction}
\end{eqnarray}
where the spin connection is defined as $\omega_M= \frac{1}{4}
\omega_M^{\bar{M} \bar{N}} \Gamma_{\bar{M}} \Gamma_{\bar{N}}$ and
\begin{eqnarray}
 \omega_M ^{\bar{M} \bar{N}}
   &=& \frac{1}{2} {e}^{N \bar{M}}(\partial_M e_N^{~\bar{N}}
                      - \partial_N e_M^{~\bar{N}}) \nn \\
   &-& \frac{1}{2} {e}^{N\bar{N}}(\partial_M e_N^{~\bar{M}}
                      - \partial_N e_M^{~\bar{M}})  \nn \\
   &-& \frac{1}{2} {e}^{P \bar{M}} {e}^{Q \bar{N}} (\partial_P e_{Q
{\bar{R}}} - \partial_Q e_{P {\bar{R}}}) {e}_M^{~\bar{R}}.
\end{eqnarray}
The non-vanishing components of the spin connection $\omega_M$ for
the background metric (\ref{linee}) are
\begin{eqnarray}
  \omega_\mu =\frac{1}{2}(\partial_{z}A) \gamma_\mu \gamma_5
             +\hat{\omega}_\mu, \label{spinConnection}
\end{eqnarray}
where $\mu=0,1,2,3$ and $\hat{\omega}_\mu=\frac{1}{4}
\bar\omega_\mu^{\bar{\mu} \bar{\nu}} \Gamma_{\bar{\mu}}
\Gamma_{\bar{\nu}}$ is the spin connection derived from the metric
$\hat{g}_{\mu\nu}(x)=\hat{e}_{\mu}^{~\bar{\mu}}
\hat{e}_{\nu}^{~\bar{\nu}}\eta_{\bar{\mu}\bar{\nu}}$. Then the
equation of motion is given by
\begin{eqnarray}
 \left\{ \gamma^{\mu}(\partial_{\mu}+\hat{\omega}_\mu)
         + \gamma^5 \left(\partial_z  +2 \partial_{z} A \right)
         -\eta\; \text{e}^A F(\phi)
 \right \} \Psi =0, \label{DiracEq1}
\end{eqnarray}
where $\gamma^{\mu}(\partial_{\mu}+\hat{\omega}_\mu)$ is the Dirac
operator on the brane.

Now we study the above 5-dimensional Dirac equation, and write the
spinor in terms of 4-dimensional effective fields. Because of the
Dirac structure of the fifth gamma matrix $\gamma^5$, we expect
that the left- and right-handed projections of the four
dimensional part to behave differently. From the equation of
motion (\ref{DiracEq1}), we will search for the solutions of the
general chiral decomposition
\begin{equation}
 \Psi(x,z) = \text{e}^{-2A}\left(\sum_n\psi_{Ln}(x) \alpha_{Ln}(z)
 +\sum_n\psi_{Rn}(x) \alpha_{Rn}(z)\right)
\end{equation}
with $\psi_{Ln}(x)=-\gamma^5 \psi_{Ln}(x)$ and
$\psi_{Rn}(x)=\gamma^5 \psi_{Rn}(x)$ the left-handed and
right-handed components of a 4D Dirac field. Here, to obtain the
equations for the basis functions $\psi_{Ln}(x)$ and
$\psi_{Rn}(x)$, we assume that $\psi_{L}(x)$ and $\psi_{R}(x)$
satisfy the 4D massive Dirac equations
$\gamma^{\mu}(\partial_{\mu}+\hat{\omega}_\mu)\psi_{Ln}(x)
=\mu_n\psi_{R_n}(x)$ and
$\gamma^{\mu}(\partial_{\mu}+\hat{\omega}_\mu)\psi_{Rn}(x)
=\mu_n\psi_{L_n}(x)$. Then the KK modes $\alpha_{Ln}(z)$ and
$\alpha_{Rn}(z)$ satisfy the following coupled equations
\begin{subequations}
\begin{eqnarray}
 \left[\partial_z
                  + \eta\;\text{e}^A F(\phi) \right]\alpha_{Ln}(z)
  &=&  ~~\mu_n \alpha_{Rn}(z), \label{CoupleEq1a}  \\
 \left[\partial_z
                  - \eta\;\text{e}^A F(\phi) \right]\alpha_{Rn}(z)
  &=&  -\mu_n \alpha_{Ln}(z), \label{CoupleEq1b}
\end{eqnarray}\label{CoupleEq1}
\end{subequations}
i.e.,
\begin{subequations}
\begin{eqnarray}
 \left[\partial_z-\eta\;\text{e}^A F(\phi)\right]
 \left[\partial_z+\eta\;\text{e}^A F(\phi) \right]
 \alpha_{Ln}(z) &=& -\mu_n^2 \alpha_{Ln}(z), \label{LeftEqa}  \\
 \left[\partial_z+\eta\;\text{e}^A F(\phi)\right]
 \left[\partial_z-\eta\;\text{e}^A F(\phi) \right]
 \alpha_{Rn}(z) &=& -\mu_n^2 \alpha_{Rn}(z). \label{LeftEqb}
\end{eqnarray}\label{CoupleEq1}
\end{subequations}
Hence, we get the Schr\"{o}dinger-like equations for the left and
right chiral fermions
\begin{eqnarray}
  \big(-\partial^2_z + V_L(z) \big)\alpha_{Ln}
            &=&\mu_n^2 \alpha_{Ln},~~
   \label{SchEqLeftFermion}  \\
  \big(-\partial^2_z + V_R(z) \big)\alpha_{Rn}
            &=&\mu_n^2 \alpha_{Rn},
   \label{SchEqRightFermion}
\end{eqnarray}
where the mass-independent potentials are given by
\begin{subequations}
\begin{eqnarray}
  V_L(z)&=& \text{e}^{2A} \eta^2 F^2(\phi)
     - \text{e}^{A} \eta\; \partial_z F(\phi)
     - (\partial_{z}A) \text{e}^{A} \eta F(\phi), \label{VL}\\
  V_R(z)&=&   V_L(z)|_{\eta \rightarrow -\eta}. \label{VR}
\end{eqnarray}\label{Vfermion}
\end{subequations}

In order to obtain the standard 4D action for the massive chiral
fermions:
\begin{eqnarray}
 S_{1/2} &=& \int d^5 x \sqrt{-g} ~\bar{\Psi}
     \left(  \Gamma^M (\partial_M+\omega_M)
     -\eta F(\phi)\right) \Psi  \nn \\
  &=& \sum_{n}\int d^4 x \sqrt{-\hat{g}}
    \left\{~\bar{\psi}_{Rn}
      \gamma^{\mu}(\partial_{\mu}+\hat{\omega}_\mu)\psi_{Rn}
        -~\bar{\psi}_{Rn}\mu_{n}\psi_{Ln} \right \}  \nn \\
    &+&\sum_{n}\int d^4 x \sqrt{-\hat{g}}
    \left\{~\bar{\psi}_{Ln}
      \gamma^{\mu}(\partial_{\mu}+\hat{\omega}_\mu)\psi_{Ln}
        -~\bar{\psi}_{Ln}\mu_{n}\psi_{Rn} \right \}  \nn \\
  &=&\sum_{n}\int d^4 x \sqrt{-\hat{g}}
    ~\bar{\psi}_{n}
      [\gamma^{\mu}(\partial_{\mu}+\hat{\omega}_\mu)
        -\mu_{n}]\psi_{n},
\end{eqnarray}
we need the following orthonormality conditions for
$\alpha_{L_{n}}$ and $\alpha_{R_{n}}$:
\begin{eqnarray}
 \int_{-\infty}^{\infty} \alpha_{Lm} \alpha_{Ln}dz
   &=& \delta_{mn}, \label{orthonormalityFermionL} \\
 \int_{-\infty}^{\infty} \alpha_{Rm} \alpha_{Rn}dz
   &=& \delta_{mn}, \label{orthonormalityFermionR}\\
 \int_{-\infty}^{\infty} \alpha_{Lm} \alpha_{Rn}dz
   &=& 0. \label{orthonormalityFermionR}
\end{eqnarray}

It can be seen that, for the left (right) chiral fermion
localization, there must be some kind of scalar-fermion coupling.
This situation can be compared with the one in the RS framework
\cite{BajcPLB2000}, where additional localization method
\cite{JackiwPRD1976} was introduced for spin 1/2 fields.
Furthermore, $F(\phi(z))$ must be an odd function of $\phi(z)$
when we demand that $V_L(z)$ or $V_R(z)$ is $Z_2$-even with
respect to the extra dimension $z$. In this paper, we will
consider two cases $F(\phi)=\phi$ and $F(\phi)=\sin(
\frac{\phi}{\phi_0})\cos^{-\delta}(\frac{\phi}{\phi_0})$ as
examples. For $F(\phi)=\phi$, we get a continuous spectrum of KK
modes with $\mu^2\geq0$. However, it is shown that even the
massless left and right chiral modes can not be localized on the
brane. For $F(\phi) = \sin(\frac{\phi}{\phi_0}) \cos^{-\delta}
(\frac{\phi}{\phi_0})$, there exists a mass gap, and we get some
discrete bound modes and a continuous spectrum of KK modes.

\subsubsection{Case I}

For the first case $F(\phi)=\phi$, the explicit forms of the
potentials (\ref{Vfermion}) are
\begin{eqnarray}
 V^S_L(z)
  &=& \eta^2 \phi_0^2 \cosh^{-2\delta} \left( \frac{\beta z}{\delta}  \right)
      \arctan^2 \sinh \left( \frac{\beta z}{\delta} \right)
      \nonumber \\
  &&  +\frac{\eta\beta\phi_0}{\delta} \cosh^{-1-\delta} \left( \frac{\beta z}{\delta}\right)
      \left[\delta \sinh \left( \frac{\beta z}{\delta} \right)
            \arctan \sinh \left( \frac{\beta z}{\delta} \right)-1
      \right],  \label{VSL_phi} \\
  V^S_R(z) &=& V^S_L(z)|_{\eta \rightarrow -\eta},
\end{eqnarray}
and
\begin{eqnarray}
 V^A_L(z)
  &=& \frac{\eta\beta^2\phi_0 \cosh^{-3/2}(2\beta z)}
           {\big(\beta+a\arctan^2\tanh(\beta z)\big)^2}
      \bigg[ a\arctan\sinh(2\beta z)  \nonumber \\
  &&       + \eta\phi_0 \sqrt{\cosh(2\beta z)}\arctan^2\sinh(2\beta z)
            \quad  \quad \quad  \quad \quad \quad \quad
            (\delta=\frac{1}{2})  \label{VAL_phi}\\
  &&       + \big(\beta+a\arctan\tanh(\beta z)\big)
             \big(\sinh(2\beta z)\arctan\sinh(2\beta z)-2\big)
      \bigg],    \nonumber \\
  V^A_R(z) &=& V^A_L(z)|_{\eta \rightarrow -\eta},
\end{eqnarray}
for the symmetric and asymmetric $dS$ brane world solutions,
respectively.

\begin{figure}[htb]
\begin{center}
\includegraphics[width=7cm,height=5cm]{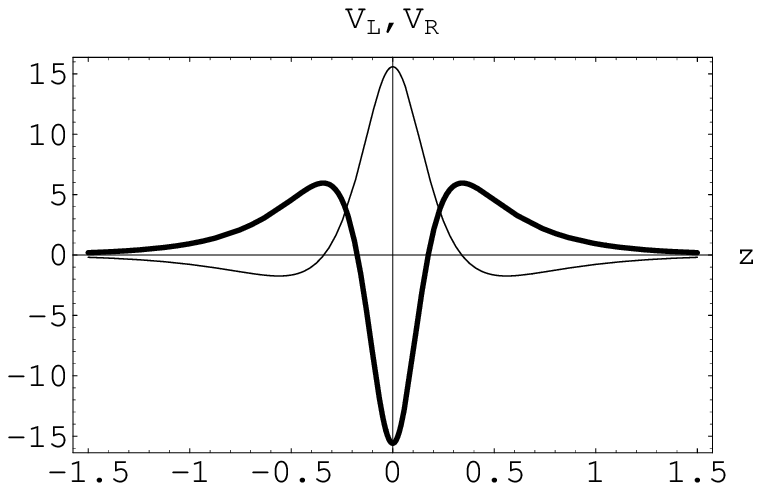}
\includegraphics[width=7cm,height=5cm]{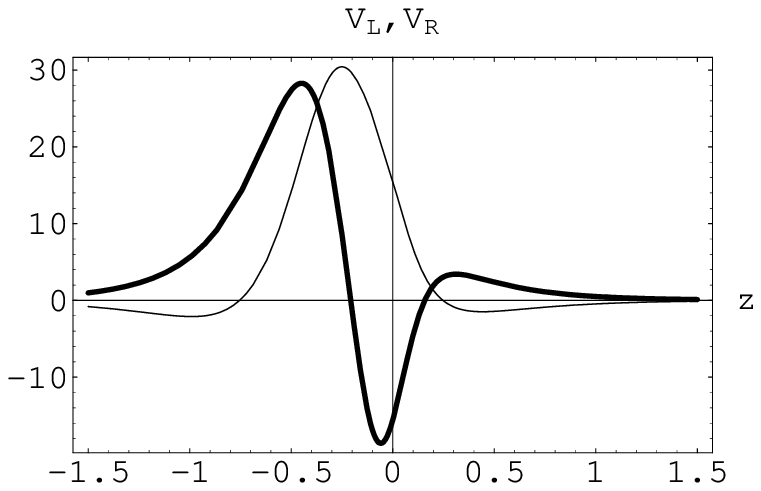}
\end{center}
\caption{The shapes of the potentials $V_L$(thick lines),
$V_R$(thin lines) for left and right chiral fermions for the case
$F(\phi)=\phi$. The parameters are set to $\delta=1/2, \eta=3,
\beta=3$, $a=0$ (left) and $a=3$ (right).}
 \label{fig_VFermion_Case1}
\end{figure}

All potentials have the asymptotic behavior:
$V_{L,R}(z\rightarrow\pm \infty)\rightarrow0$. But for a given
coupling constant $\eta$ and the parameter $\beta$, the values of
the potentials for the left and right chiral fermions at $z=0$ are
opposite. The shapes of the potentials are shown in Fig.
\ref{fig_VFermion_Case1} for given values of positive $\eta$ and
$\beta$. It can be seen that $V_L(z)$ is indeed a modified volcano
type potential. Hence, the potential provides no mass gap to
separate the fermion zero mode from the excited KK modes, and
there exists a continuous gapless spectrum of the KK modes for
both the left chiral and right chiral fermions.

For positive $\beta$ and $\eta$, only the potential for left
chiral fermions has a negative value at the location of the brane,
which could trap the left chiral fermion zero mode solved from
(\ref{CoupleEq1a}) by setting $\mu_0=0$:
\begin{equation}
 {\alpha}_{L0}(z)
 \propto \exp\left(-\eta\int^z dz'\text{e}^{A(z')}\phi(z')\right).
  \label{zeroMode1}
\end{equation}
In order to check the normalization condition
(\ref{orthonormalityFermionL}) for the zero mode
(\ref{zeroMode1}), we need to check whether the inequality
\begin{equation}
\int dz \exp\left(-2\eta\int^z dz'
  \text{e}^{A(z')}\phi(z')\right)
  < \infty     \label{condition1}
\end{equation}
is satisfied. For the integral $\int dz\text{e}^{A}\phi$, we only
need to consider the asymptotic characteristic of the function
$\text{e}^{A}\phi$ for $z \rightarrow \infty$. For symmetric $dS$
brane case, noting that $\arctan(\sinh z) \rightarrow \pi/2$ when
$z \rightarrow \infty$, we have
\begin{eqnarray}
 &&\text{e}^{A}\phi
  =\phi_{0}\cosh^{-\delta}\left(\frac{\beta z}{\delta}\right)
   \arctan\left(\sinh \frac{\beta  z}{\delta}\right)
   \rightarrow
   \frac{\pi}{2}\phi_{0} 2^{\delta}  \text{e}^{-\beta z} ,~~~ \\
 && \exp\left(-2\eta \int dz\text{e}^{A}\phi\right)
    \rightarrow
    \exp\left(2^{\delta}\pi\eta \phi_{0}   \text{e}^{-\beta z}/\beta\right)
    \rightarrow 1,
\end{eqnarray}
which indicates that the normalization condition
(\ref{condition1}) is not satisfied and the zero mode of the left
chiral fermions can not be localized on the brane. For asymmetric
case, we can also get the same conclusion. This is different from
the conclusion obtained in Refs. \cite{KoleyCQG2005,LiuPRD2008},
where the zero mode of the left chiral fermions can be localized
on the Branes in the Background of Sine-Gordon Kinks.

\subsubsection{Case II}

For the case $F(\phi)=\sin(
\frac{\phi}{\phi_0})\cos^{-\delta}(\frac{\phi}{\phi_0})$, the
potentials (\ref{Vfermion}) are
\begin{eqnarray}
 V^S_L(z)
  &=& \eta\bigg(\eta-\frac{\beta+\delta\eta}{\delta}
                     \text{sech}^{2}(\beta z/\delta) \bigg),
      \label{VSL_II} \\
  V^S_R(z) &=& V^S_L(z)|_{\eta \rightarrow -\eta},\label{VSR_II}
\end{eqnarray}
and
\begin{eqnarray}
 V^A_L(z)
  &=& \frac{\eta\beta^2\text{sech}(2\beta z)\tanh(2\beta z)
               \big(a+\eta\sinh(2\beta z) \big)}
           {(\beta+a\arctan\tanh(\beta z))^2}
     -\frac{2\eta\beta^2\text{sech}^2(2\beta z)}
           {\beta+a\arctan\tanh(\beta z)},  \label{VAL_II}\\
  V^A_R(z) &=& V^A_L(z)|_{\eta \rightarrow -\eta},\label{VAR_II}
\end{eqnarray}
for the symmetric and asymmetric $dS$ brane world solutions,
respectively.\\

\noindent {\textbf{1. Symmetric $dS$ brane}}\\

\begin{figure}[htb]
\begin{center}
 \includegraphics[width=7cm,height=5cm]{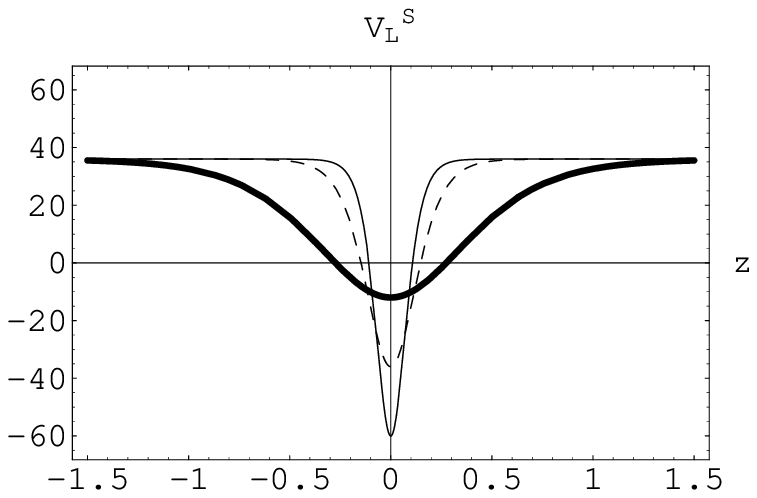}
 \includegraphics[width=7cm,height=5cm]{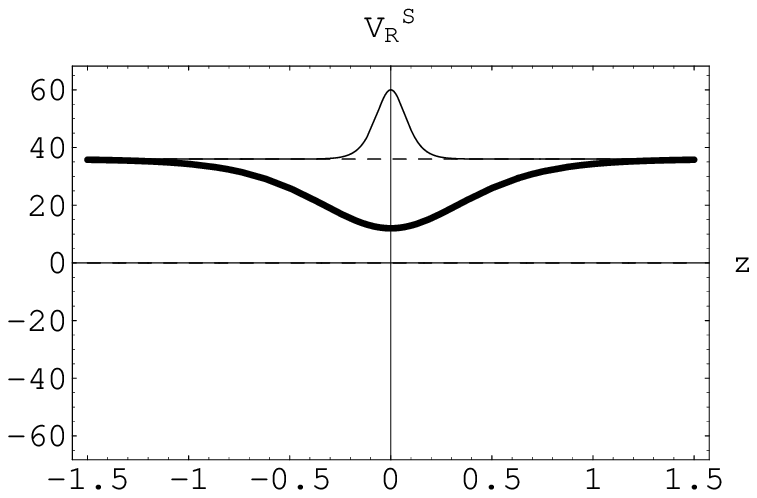}
\end{center} \caption{The shapes of the potentials $V^S_L$ (left)
and $V^S_R$ (right) of left and right chiral fermions for the
symmetric $dS$ brane with different $\beta$. The parameters are
set to $\delta=1/2$, $\eta=6$, and $\beta=1$ for thick lines,
$\beta=3$ for dashed lines and $\beta=5$ for thin lines.}
 \label{fig_VsLR_beta}
\end{figure}

\begin{figure}[htb]
\begin{center}
\includegraphics[width=7cm,height=5cm]{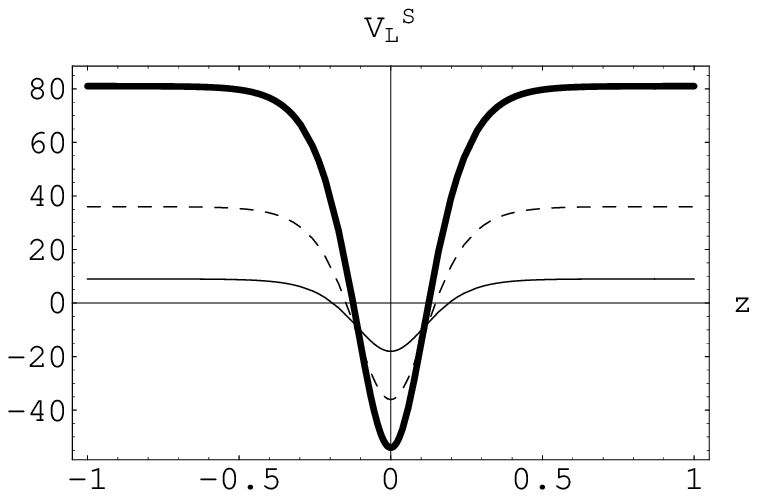}
\includegraphics[width=7cm,height=5cm]{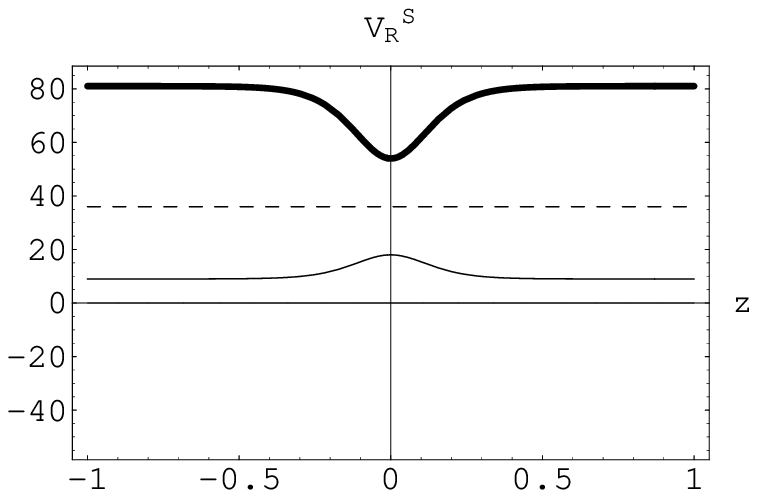}
\end{center} \caption{The shapes of the potentials $V^S_L$ (left)
and $V^S_R$ (right) of left and right chiral fermions for the
symmetric $dS$ brane with different $\eta$. The parameters are set
to $\delta=1/2$, $\beta=3$, and $\eta=9$ for thick lines, $\eta=6$
for dashed lines and $\eta=3$ for thin lines.}
 \label{fig_VsLR_eta}
\end{figure}

We first investigate the case of symmetric $dS$ brane. The values of
the corresponding potentials (\ref{VSL_II}) and (\ref{VSR_II}) at $y
= 0$ and $y = \pm\infty$ are given by
\begin{eqnarray}
V^S_L(0) ~&=&-V^S_R(0) = -\frac{\beta\eta}{\delta}, \\
V^S_L(\pm\infty) &=&V^S_R(\pm\infty) = \eta^2,
\end{eqnarray}
i.e., both potentials have same asymptotic behavior when $y
\rightarrow \pm\infty$, but opposite behavior at the origin $z=0$.
The shapes of the two potentials are shown in Figs.
\ref{fig_VsLR_beta} and \ref{fig_VsLR_eta} for different values of
$\beta$ and $\eta$, respectively.

Note that, for a positive coupling constant $\eta$, the potential
for left chiral fermions has a negative value at the location of
the brane and a positive value far away from the brane along the
extra dimension, which can always trap the left chiral fermion
zero mode:
\begin{equation}
 {\alpha}^S_{L0} = \left[
  \frac{\beta\;\Gamma(\frac{\beta+\delta\eta}{2\beta})}
       {\delta\sqrt{\pi}\;\Gamma(\frac{\delta\eta}{2\beta})}\right]^{\frac{1}{2}}
  \cosh^{-{\delta\eta}/{\beta}}(\beta z/\delta).     ~~~~~(\eta>0)
  \label{zeroModeLeftFermion}
\end{equation}
The zero mode represents the lowest energy eigenfunction of the
Schr\"{o}dinger equation (\ref{SchEqLeftFermion}) since it has no
nodes. The right chiral fermion zero mode for $\eta>0$ is not
localized on the brane, which can be seen from the potential
$V^S_R$ in Figs. (\ref{fig_VsLR_beta}) and (\ref{fig_VsLR_eta}).

It is clear that for $\mu^2_{Ln} > \eta^2$, we obtain the
asymptotic plane waves. The general bound states for the potential
$V^S_L$ (\ref{VSL_II}) for left chiral fermions are found to be
\begin{eqnarray}
  {\alpha}^S_{Ln}  \propto
  \cosh^{1+\frac{\delta\eta}{\beta}} \left(\frac{\beta z}{\delta}\right)~
  {_2}F_1 \left(a_n,b_n;\frac12;-\sinh^2(\beta z/\delta)\right),
\end{eqnarray}
for even $n$ and
\begin{eqnarray}
  {\alpha}^S_{Ln}  \propto
   \cosh^{1+\frac{\delta\eta}{\beta}} \left(\frac{\beta z}{\delta}\right)
   \sinh\left(\frac{\beta z}{\delta}\right) ~
  {_2}F_1\left(a_n+\frac12,b_n+\frac12;\frac32;
                  -\sinh^2({\beta z}/{\delta})\right),
\end{eqnarray}
for odd $n$, where ${_2}F_1$ is the hypergeometric function, the
parameters $a_n$ and $b_n$ are given by
\begin{eqnarray}
 a_n = \frac12 \left(n+1\right), \quad
 b_n = \frac{\delta\eta}{\beta} -\frac12 \left(n-1\right).
\end{eqnarray}
The corresponding mass spectrum of the bound states is
\begin{eqnarray}
 \mu^2_{Ln} = \frac{\beta(2\delta\eta-\beta n)n}{\delta^2},~~~
 (\eta>0,~n=0,1,2,...<\frac{\delta\eta}{\beta}).~~
 \label{massSpectrumVL}
\end{eqnarray}
It shows that the ground state always belongs to the spectrum of
$V^S_L(z)$ for positive $\eta$, which is just the zero mode
(\ref{zeroModeLeftFermion}) with $\mu_{L0}=0$. Since the ground
state has the lowest mass square $\mu^2_{L0}=0$, there is no
tachyonic left chiral fermion modes. Here, we suppose the number of
bound states for left chiral fermions is $N_L$. If
$0<\eta\leq\beta/\delta$, there is only one bound state ($N_L=1$),
i.e., the zero mode (\ref{zeroModeLeftFermion}). In order to get
bound exited states ($N_L\geq 2$), we need the condition
$\eta>\beta/\delta$.

\begin{figure}[htb]
 \centering
 \subfigure[Left chiral fermions, $a=0$]{\label{fig_FermionMn2VaveVLa_0}
  \includegraphics[width=6.5cm,height=4.5cm]{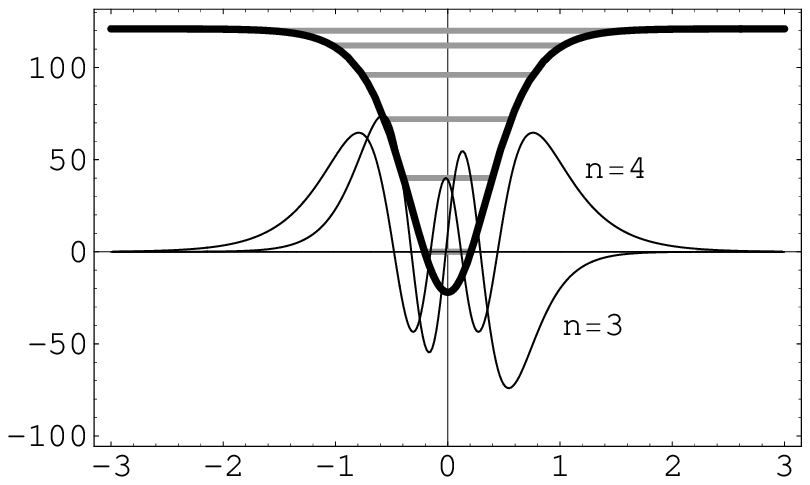}}
 \subfigure[Right chiral fermions, $a=0$]{\label{fig_FermionMn2VaveVRa_0}
  \includegraphics[width=6.5cm,height=4.5cm]{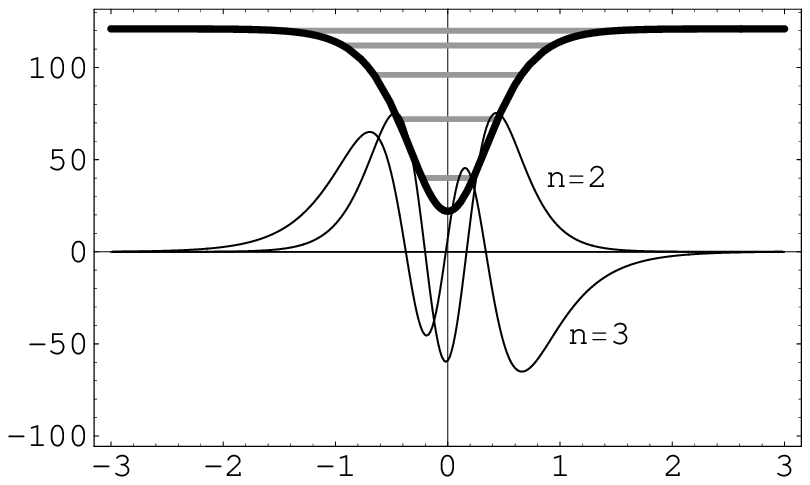}}
 \subfigure[Left chiral fermions, $a=0.1$]{\label{fig_FermionMn2VaveVLa_0.1}
  \includegraphics[width=6.5cm,height=4.5cm]{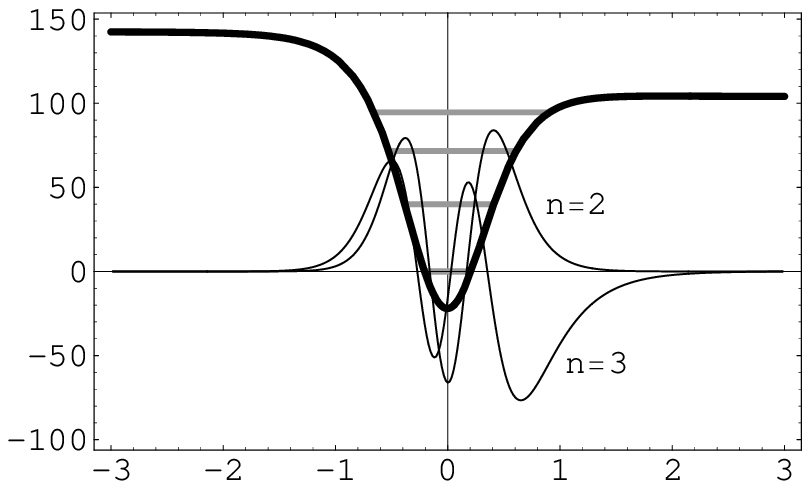}}
 \subfigure[Right chiral fermions, $a=0.1$]{\label{fig_FermionMn2VaveVRa_0.1}
  \includegraphics[width=6.5cm,height=4.5cm]{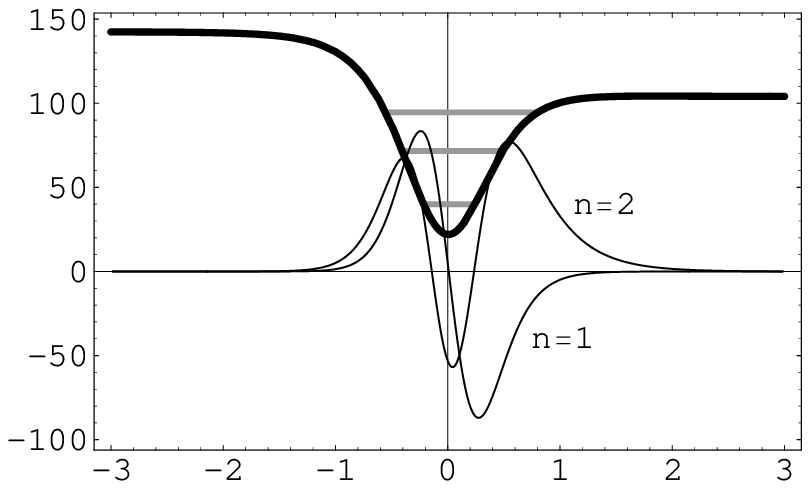}}
 \subfigure[Left chiral fermions, $a=0.5$]{\label{fig_FermionMn2VaveVLa_0.5}
  \includegraphics[width=6.5cm,height=4.5cm]{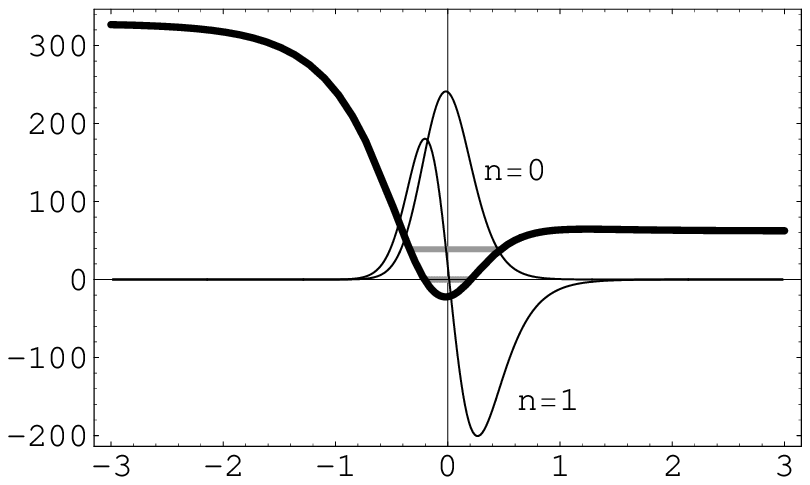}}
 \subfigure[Right chiral fermions, $a=0.5$]{\label{fig_FermionMn2VaveVRa_0.5}
  \includegraphics[width=6.5cm,height=4.5cm]{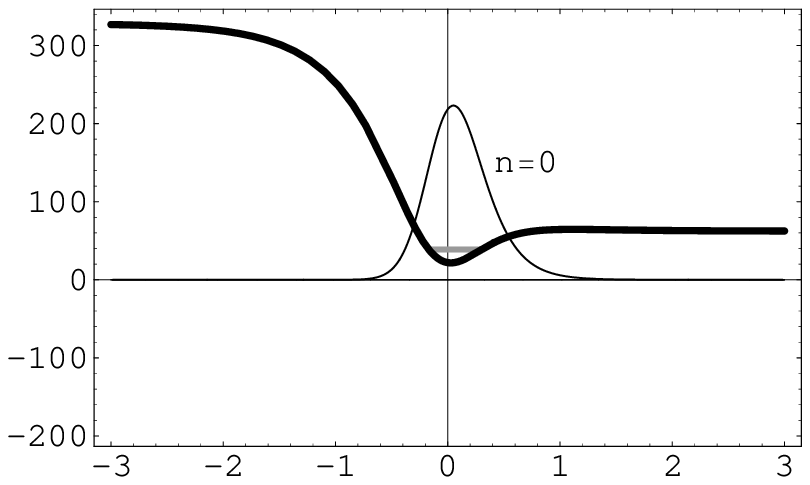}}
\caption{The potentials $V^A_{L,R}(z)$ (black thick lines), the
mass spectrum $\mu^2_{L,R}$ (thick gray lines) and some
eigenfunctions (black thin lines) for asymmetric $dS$ brane with
$\beta=1$, $\delta=1/2$, $\eta=11$ and different $a$.}
 \label{fig_FermionMn2VaveV}
\end{figure}

In the case $\eta>0$, the potential $V^S_R(z)=
\eta\big(\eta+\frac{\beta-\delta\eta}{\delta} \text{sech}^{2}(\beta
z/\delta) \big)$ for right chiral fermions is always positive near
the location of the brane, which shows that it can not trap the
right chiral zero mode. For the case $0<\eta<\beta/\delta$, we have
$V^S_R(0)\geq V^S_R(\pm\infty)>0$, which shows that there is no any
bound state for the potential of right chiral fermions. For the
special value $\eta=\beta/\delta$, the potential $V^S_R$ is a
positive constant: $V^S_R(z)=\eta^2=\beta^2/\delta^2$, and there is
still no any bound state. However, provided $\eta> \beta/\delta$, we
will get a potential well since $V^S_R(0)< V^S_R(\pm\infty)$ (see
Figs. \ref{fig_VsLR_beta} and \ref{fig_VsLR_eta}), which indicates
that there may be some bound states, but none of them is zero mode.
The general bound states for the potential $V^S_R$ are
\begin{eqnarray}
  {\alpha}^S_{Rn}  \propto
  \cosh^{\frac{\delta\eta}{\beta}} \left(\frac{\beta z}{\delta}\right)~
  {_2}F_1 \left(\frac{1+n}{2},
                \frac{\delta\eta}{\beta}-\frac{1+n}{2};
                \frac12;
                -\sinh^2(\beta z/\delta)\right)
\end{eqnarray}
for even $n$ and
\begin{eqnarray}
  {\alpha}^S_{Rn}  \propto
   \cosh^{\frac{\delta\eta}{\beta}} \left(\frac{\beta z}{\delta}\right)
   \sinh\left(\frac{\beta z}{\delta}\right) ~
  {_2}F_1\left(1+\frac{n}{2},
               \frac{\delta\eta}{\beta}-\frac{n}{2};
               \frac32;
               -\sinh^2({\beta z}/{\delta})\right)
\end{eqnarray}
for odd $n$. The corresponding mass spectrum is
\begin{eqnarray}
 \mu^2_{Rn} = \frac{(n+1)\beta(2\delta\eta- (n+1)\beta)}{\delta^2},~~~
   (\eta> \frac{\beta}{\delta},~
    n=0,1,2,...<\frac{\delta\eta}{\beta}-1).~~
 \label{massSpectrumVR}
\end{eqnarray}
By comparing with the mass spectrum of left chiral fermions
(\ref{massSpectrumVR}), we come to the conclusion that the number
of bound states of right chiral fermions $N_R$ is one less than
that of left ones, i.e., $N_R=N_L-1$. If $0<\eta\leq
\beta/\delta$, there is only one left chiral fermion bound state
(the zero mode). If $\eta>\beta/\delta$, there are $N_L(N_L\geq
2)$ left chiral fermion bound states and $N_L-1$ right chiral
fermion bound states. The ground state for right chiral fermions
is
\begin{equation}
 \alpha^S_{R0} = \left[
  \frac{\beta\;\Gamma(\frac{\delta\eta}{2\beta})}
       {\delta\sqrt{\pi}\;\Gamma(\frac{\delta\eta}{2\beta}-\frac{1}{2})}
       \right]^{\frac{1}{2}}
  \cosh^{1-\frac{\delta\eta}{\beta}}
  \left(\frac{\beta z}{\delta}\right), ~~~~
            \left(\eta>\frac{\beta}{\delta}\right)
  \label{RightFermionGroundState}
\end{equation}
which is not zero mode any more because the mass is determined by
$\mu^2_{R0}=\beta(2\delta\eta-
\beta)/{\delta^2}>{\beta^2}/{\delta^2}>0$. In Figs.
\ref{fig_FermionMn2VaveVLa_0} and \ref{fig_FermionMn2VaveVRa_0} we
plot the potentials, the mass spectra and some bound states of
left and right chiral fermions. For the case $\delta=1/2, \beta=1,
\eta=11$, there are 6 and 5 bound states for the left and the
right chiral fermions respectively and the mass spectra are
\begin{eqnarray}
 \mu_{Ln}^2&=&\{0, 40, 72, 96, 112, 120\} \cup [121,\infty), \\
 \mu_{Rn}^2&=&\{~~~40, 72, 96, 112, 120\} \cup [121,\infty).
\end{eqnarray}

\noindent {\textbf{2. Asymmetric $dS$ brane}}

\begin{figure}[htb]
\begin{center}
 \includegraphics[width=7cm,height=5cm]{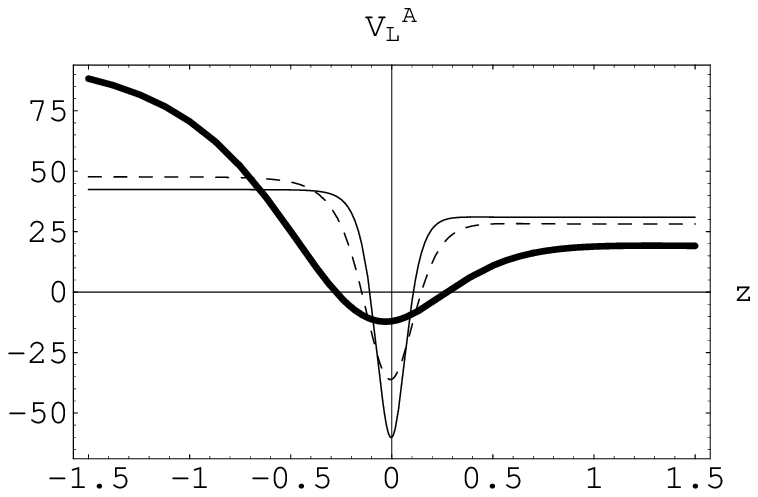}
 \includegraphics[width=7cm,height=5cm]{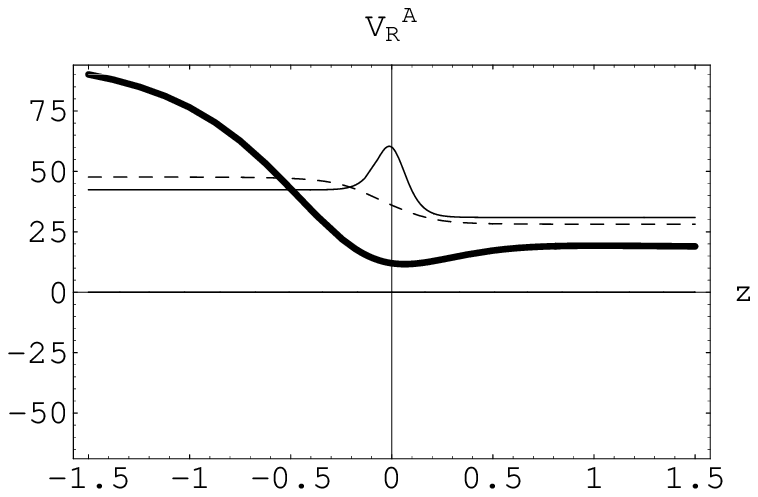}
\end{center} \caption{The shapes of the potentials $V^A_L$ (left)
and $V^A_R$ (right) of left and right chiral fermions for the
asymmetric $dS$ brane with different $\beta$. The parameters are
set to $\delta=1/2$, $\eta=6$, $a=0.5$, and $\beta=1$ for thick
lines, $\beta=3$ for dashed lines and $\beta=5$ for thin lines.}
 \label{fig_VaLR_beta}
\end{figure}

\begin{figure}[htb]
\begin{center}
\includegraphics[width=7cm,height=5cm]{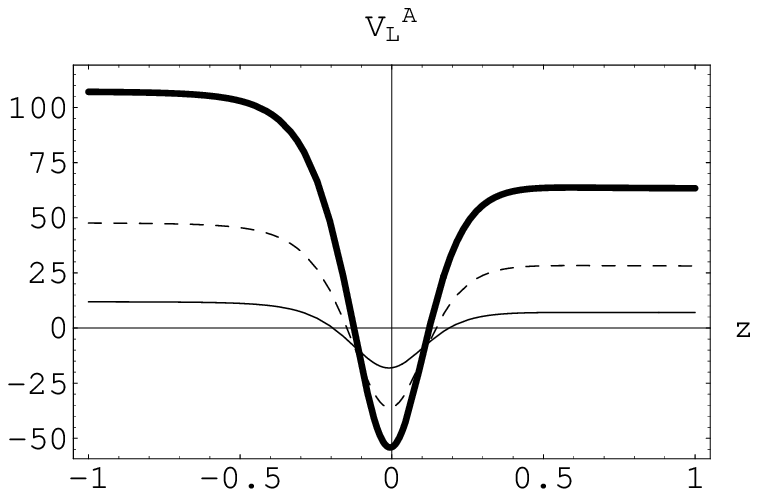}
\includegraphics[width=7cm,height=5cm]{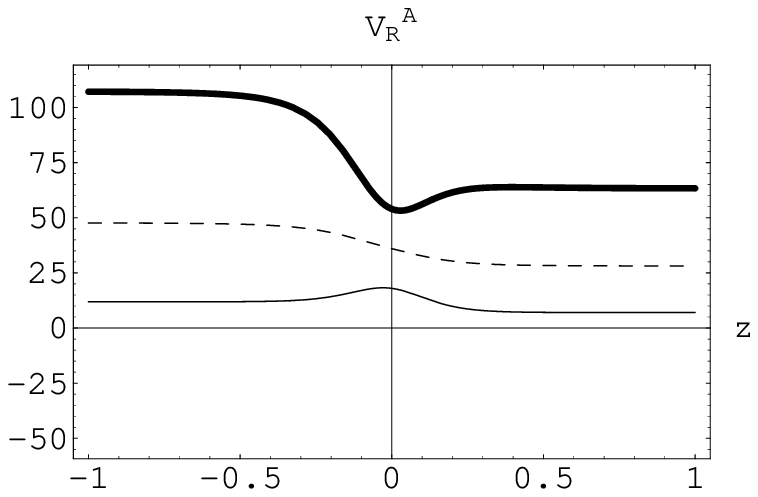}
\end{center} \caption{The shapes of the potentials $V^A_L$ (left)
and $V^A_R$ (right) of left and right chiral fermions for the
asymmetric $dS$ brane with different $\eta$. The parameters are
set to $\delta=1/2$, $\beta=3$, $a=0.5$, and $\eta=9$ for thick
lines, $\eta=6$ for dashed lines and $\eta=3$ for thin lines.}
 \label{fig_VaLR_eta}
\end{figure}

Now we turn to the case of asymmetric $dS$ brane, for which the
corresponding potentials (\ref{VAL_II}) and (\ref{VAR_II}) are
obviously asymmetric and the solution of the bound states and mass
spectrum is very complex. The values of the potentials at $z = 0,
\pm\infty$ are given by
\begin{eqnarray}
V^A_L(0) ~&=&-V^A_R(0) = -{2\beta\eta}, \label{VALR0} \nonumber \\
V^A_L(+\infty) &=&V^A_R(+\infty) =
\frac{16\beta^2\eta^2}{(a\pi+4\beta)^2}<\eta^2, \label{VALRrightLimit}\\
V^A_L(-\infty) &=&V^A_R(-\infty) =
\frac{16\beta^2\eta^2}{(a\pi-4\beta)^2}>\eta^2.
\label{VALRleftLimit}  \nonumber
\end{eqnarray}
Both potentials have also same asymptotic behavior when $z
\rightarrow \pm\infty$. However, compared with the symmetric
potentials (\ref{VSL_II}) and (\ref{VSR_II}), $V^A_{L,R}(-\infty)$
increase and $V^A_{L,R}(+\infty)$ decrease for positive asymmetric
factor $a$, which may reduce the number of the bound states. The
shapes of the two potentials are shown in Figs.
\ref{fig_VaLR_beta} and \ref{fig_VaLR_eta} for different values of
$\beta$ and $\eta$, respectively. Different from the symmetric
potentials $V^S_{L,R}(z)$, the asymmetric ones $V^A_{L,R}(z)$ at
$z=\pm\infty$ dependent on the parameter $\beta$ unless the
asymmetric factor $a=0$. Hence, even at same $\eta$ and $a$, the
limits of $V^A_{L,R}(z)$ at $z\rightarrow +\infty$ and
$z\rightarrow -\infty$ are different for different $\beta$ (see
Fig. \ref{fig_VaLR_beta}). For positive $\eta$, the right chiral
fermion zero mode does not exist, but the left one is always exist
and can be solved as
\begin{equation}
 {\alpha}^A_{L0}(z)
 \propto \exp\left(-\eta\beta\int^z dz'
         \frac{\tanh(2\beta z')}{\beta+a\arctan\tanh(\beta z')}\right).
  \label{FermionZeroModeLeftA}
\end{equation}

For $\mu^2_{n} > {16\beta^2\eta^2}/{(a\pi+4\beta)^2}$, we obtain
the continuum of asymptotic plane waves. In order to obtain
acceptable normalizable modes, $\mu^2_n$ should be limited in the
interval $[0,{16\beta^2\eta^2}/{(a\pi+4\beta)^2})$. Although the
analytic massive modes can not be solved because of the complexity
of the potentials, we can get the numerical solutions. The mass
spectra are listed in Tab. \ref{TabMassSpectrumL} for left chiral
fermions and Tab. \ref{TabMassSpectrumR} for right ones for some
given parameters. We also plot the potentials, mass spectra and
part of the eigenfunctions in Fig. \ref{fig_FermionMn2VaveV}. From
these tables and Eq. (\ref{VALRrightLimit}), we can draw a
conclusion: the number of the bound states increases with the
coupling constant $\eta$ but decreases with the asymmetric factor
$a$.

\begin{table}[h]
\begin{center}
\begin{tabular}{|c|c|r|r|l|}
 \hline
  $a$ & $N_L$ & $V^A_{L}(+\infty)$ & $V^A_{L}(-\infty)$ & Mass spectrum $\mu^2_{Ln}$ of bound states \\
 \hline \hline
  0.00 & 6 & 121.0~~ &    121.0 & \{0, ~40.00, ~72.00, ~96.00, ~112.00, ~120.00\} \\
  0.02 & 5 & 117.3~~ &    124.9 & \{0, ~40.00, ~71.99, ~95.94, ~111.74\} \\
  0.10 & 4 & 104.0~~ &    142.5 & \{0, ~39.95, ~71.67, ~94.57\} \\
  0.25 & 3 &  84.5~~ &    187.4 & \{0, ~39.70, ~69.95\} \\
  0.50 & 2 &  62.3~~ &    328.1 & \{0, ~38.79\} \\
  1.25 & 1 &  30.8~~ & 363203.8  & \{0 \} \\
  \hline
\end{tabular}\\
\caption{Mass spectrum of bound states for asymmetric potentials
$V^A_{L}(z)$. The parameters are set to $\delta=1/2$, $\beta=1$,
$\eta=11$. $N_L$ presents the number of bound states for left
chiral fermions.} \label{TabMassSpectrumL}
\end{center}
\end{table}

\begin{table}[h]
\begin{center}
\begin{tabular}{|c|c|r|r|l|}
 \hline
  $a$ & $N_R$ & $V^A_{R}(+\infty)$ & $V^A_{R}(-\infty)$ & Mass spectrum $\mu^2_{Rn}$ of bound states \\
 \hline \hline
  0.00 & 5 & 121.0~~ &    121.0 & \{40.00, ~72.00, ~96.00, ~112.00, ~120.00\} \\
  0.02 & 4 & 117.3~~ &    124.9 & \{40.00, ~71.99, ~95.94, ~111.74\} \\
  0.10 & 3 & 104.0~~ &    142.5 & \{39.95, ~71.67, ~94.57\} \\
  0.25 & 2 &  84.5~~ &    187.4 & \{39.70, ~69.95\} \\
  0.50 & 1 &  62.3~~ &    328.1 & \{38.79\} \\
  1.25 & 0 &  30.8~~ & 363203.8 & \{~ \} \\
  \hline
\end{tabular}\\
\caption{Mass spectrum of bound states for asymmetric potentials
$V^A_{R}(z)$. The parameters are set to $\delta=1/2$, $\beta=1$,
$\eta=11$. $N_R$ presents the number of bound states for right
chiral fermions.} \label{TabMassSpectrumR}
\end{center}
\end{table}

To close this section, we make some comments on the issue of the
localization of fermions. Localizing the fermions on branes or
defects requires us to introduce other interactions besides
gravity. More recently, Volkas {\em et al } had extensively
analyzed localization mechanisms on a domain wall. In particular,
in Ref. \cite{Volkas0705.1584}, they proposed a well-defined model
to localize the SM, or something close to it, on a domain wall
brane. There are some other backgrounds, for example, gauge field
\cite{LiuJHEP2007}, supergravity \cite{Mario,Parameswaran0608074}
and vortex background
\cite{LiuNPB2007,LiuVortexFermion,Rafael200803,StojkovicPRD},
could be considered. The topological vortex coupled to fermions
may result in chiral fermion zero modes \cite{JackiwRossiNPB1981}.

\section{Conclusion and discussion}

In this paper, by presenting the shapes of the mass-independent
potentials of KK modes in the corresponding Schr\"{o}dinger
equations, we have investigated the localization and mass spectra of
various matter fields with spin 0, 1 and 1/2 on symmetric and
asymmetric $dS$ thick branes, where the asymmetric $dS$ thick brane
is constructed from the symmetric one by using a same scalar (kink)
with different potentials.

For spin 0 scalars and spin 1 vectors, the potentials of KK modes
in the corresponding Schr\"{o}dinger equations are the modified
P\"{o}schl-Teller potentials. They have a finite negative well at
the location of the brane and a finite positive barrier at each
side which doesn't vanishes. Such potentials suggest that there
exist a mass gap and a series of continuous spectrum starting at
positive $\mu^2$. It can be shown that the existence of such a
mass gap is universal for all such $dS$ branes.

For the symmetric $dS$ brane, the spectrum of scalar KK modes
consists of a zero mode and a set of continuous modes, i.e., there
is only one bound mode (the zero mode). The massless mode is
separated by a mass gap from the continuous modes. For the
asymmetric $dS$ brane with a small asymmetric factor, the spectrum
is same as the symmetric case. However, for a large enough
asymmetric factor, the spectrum of scalar KK modes contains a
bound massive KK mode besides a zero mode and a set of continuous
modes, namely, there are two bound modes. For spin 1 vectors, the
spectra of KK modes on both $dS$ branes are made up of a bound
zero mode and a set of continuous ones. The asymmetric factor does
not change the number of the vector bound modes.

It is shown that, without scalar-fermion coupling, there is no bound
state for both the left and right chiral fermions. Hence, in order
to localize the massless and massive left or right chiral fermions
on the branes, some kind of kink-fermion coupling should be
introduced. As examples, two types of kink-fermion couplings are
investigated in detail. These situations can be compared with the
case of the domain wall in the RS framework \cite{BajcPLB2000},
where for localization of spin 1/2 field additional localization
method by Jackiw and Rebbi \cite{JackiwPRD1976} was introduced.

For the usual Yukawa coupling $\eta\bar{\Psi}\phi\Psi$, the
potential for only one of the left and right chiral fermions has a
finite well at the location of the brane and a finite barrier at
each side which vanishes asymptotically. It is shown that there is
only one single bound state (zero mode) which is just the lowest
energy eigenfunction of the Schr\"{o}dinger equation for the
corresponding chiral fermions. Since the potentials for both left
and right chiral fermions vanish asymptotically when far away from
the brane, all values of $\mu^2>0$ are allowed, and there exists
no mass gap but a continuous gapless spectrum of KK states with
$\mu^2>0$. The massive KK modes asymptotically turn into
continuous plane waves when far away from the brane
\cite{Lykken,dewolfe}, which represent delocalized massive KK
fermions.

For the scalar-fermion coupling $\eta\bar{\Psi}\sin(
\frac{\phi}{\phi_0})\cos^{-\delta}(\frac{\phi}{\phi_0})\Psi$ with
positive $\eta$, the potential for the left chiral fermions has a
finite well at the location of the brane, and a finite positive
barrier at each side, which does not vanishes when far away from
the brane. The potential is the modified P\"{o}schl-Teller
potential and suggest that there exist some discrete KK modes and
a series of continuous ones. The discrete modes are bound states
while the continuous ones are not. The total number of bound
states is determined by four parameters: $\delta$, $\beta$, $a$
and $\eta$. The number of bound states of right chiral fermions is
one less than that of left ones. The number of the bound states
increases with the coupling constant $\eta$. For the case of the
symmetric $dS$ brane, if $0<\eta<\beta/\delta$, there is only one
left chiral fermion bound state which is just the left chiral
fermion zero mode; if $\eta>\beta/\delta$, there are $N_L(N_L\geq
2)$ left chiral fermion bound states (including zero mode and
massive KK modes) and $N_L-1$ right chiral fermion bound states
(including only massive KK modes). For the asymmetric $dS$ brane
scenario, the asymmetric factor $a$ reduces the number of the
bound fermion KK modes. For large enough $a$, there would not be
any right chiral fermion bound mode, but at least one left chiral
fermion bound mode, i.e., the zero mode.

For fermions, localization property is decided by the coupling of
fermion and scalar. For the first type of Yukawa coupling,
$F(\phi(z)){\sim}\arctan(\sinh z)$ is a usual kink which is almost
a constant at large $z$. For the second type of coupling,
$F(\phi(z))$ is another kink likes $\sinh z$, which increases
quickly with $z$. In short, at large $z$, the first coupling is
invariant, but the second one becomes stronger. Hence, the two
different types of Yukawa couplings give different localization
properties for fermions.

Finally, we give some brief discussion about graviton and
gravitino localization on the studied branes. The Schr\"{o}dinger
potentials of graviton and gravitino KK modes are the same as that
of scalar and fermion, respectively. Thus, for the symmetric $dS$
brane and the asymmetric $dS$ brane with small asymmetric factor,
the spectrum of graviton KK modes consists of a discrete zero mode
and a set of continuous modes. While for a large enough asymmetric
factor, the spectrum of graviton KK modes contains a bound massive
KK mode besides a zero mode and a set of continuous modes. The
spectrum of gravitino is similar to that of fermion.

\section*{Acknowledgement}

The authors are really grateful to the referee for his/her
constructive comments and suggestions which considerably improved
the paper. This work was supported by the National Natural Science
Foundation of China (No. 10705013), the Doctor Education Fund of
Educational Department of China (No. 20070730055) and the
Fundamental Research Fund for Physics and Mathematics of Lanzhou
University (No. Lzu07002).

\end{document}